\documentclass[12pt,a4paper]{article}
\usepackage{a4wide}
\usepackage{amsmath}
\usepackage{amssymb}
\usepackage{epsfig}
\usepackage{subfigure}
\usepackage{exscale}
\usepackage{float}
\usepackage{bbm}
\usepackage[numbers,sort&compress]{natbib}

\newcommand{\1}{{\mathbbm{1}}}
\newcommand{\p}{\partial}

\newcommand{\ontopof}[2]{\genfrac{}{}{0pt}{}{#1}{#2}}
\newcommand{\Dag}{^{\dagger}}
\newcommand{\up}{\uparrow}
\newcommand{\down}{\downarrow}
\makeatletter
\@addtoreset{equation}{section}
\makeatother

\title{Systematic Low-Energy Effective Field Theory for Magnons and Holes in
an Antiferromagnet on the Honeycomb Lattice}

\author{F.\ K\"ampfer$^a$, B.\ Bessire$^b$, M.\ Wirz$^c$, \\
C.\ P.\ Hofmann$^d$, F.-J.\ Jiang$^e$, and U.-J.\ Wiese$^f$
\\ \\
\small $^a$ BKW FMB Energy Ltd, Energy Trading Unit, 3000 Bern, Switzerland \\
\small $^b$ Institute of Applied Physics, Bern University, CH-3012 Bern, 
Switzerland \\
\small $^c$ Mathematical Institute, Bern University, CH-3012 Bern, 
Switzerland \\
\small $^d$ Facultad de Ciencias, Universidad de Colima, Colima C.P.\ 28045, 
Mexico \\
\small $^e$ Department of Physics, National Taiwan Normal University, \\
\small 88, Sec.\ 4, Ting-Chou Rd.\, Taipei 116, Taiwan \\
\small $^f$ Albert Einstein Center for Fundamental Physics, \\
\small Institute for Theoretical Physics, Bern University, \\
\small Sidlerstrass 5, CH-3012 Bern, Switzerland}

\begin{document} 
\maketitle

\begin{abstract} \normalsize

Based on a symmetry analysis of the microscopic Hubbard and $t$-$J$ models,
a systematic low-energy effective field theory is constructed for hole-doped
antiferromagnets on the honeycomb lattice. In the antiferromagnetic phase,
doped holes are massive due to the spontaneous breakdown of the $SU(2)_s$ 
symmetry, just as nucleons in QCD pick up their mass from spontaneous chiral 
symmetry breaking. In the broken phase the effective action contains a 
single-derivative term, similar to the Shraiman-Siggia term in the square 
lattice case. Interestingly, an accidental continuous spatial rotation 
symmetry arises at leading order. As an application of the effective field
theory we consider one-magnon exchange between two holes and the formation of
two-hole bound states. As an unambiguous prediction of the effective theory, 
the wave function for the ground state of two holes bound by magnon exchange 
exhibits $f$-wave symmetry.

\end{abstract}
 
\maketitle
 
\newpage

\section{Introduction}

The physics of correlated electron systems is strongly influenced by the
geometry of the underlying crystal lattice. For example, at weak coupling the 
half-filled Hubbard model on the honeycomb lattice is a semi-metal with 
massless fermion excitations residing in two Dirac cones. This situation is 
realized in graphene. At stronger coupling the $SU(2)_s$ symmetry breaks 
spontaneously and the system becomes an antiferromagnet, which may be realized 
in the dehydrated version of Na$_2$CoO$_2 \times y$H$_2$O. On a square lattice,
on the other hand, due to Fermi surface nesting, the system is an
antiferromagnet even at arbitrarily weak coupling. Upon doping,
antiferromagnets on both the square and the  honeycomb lattice may become
high-temperature superconductors. Recenty, a spin-liquid phase was identified
in numerical simulations between the free fermion graphene phase and the 
strongly correlated antiferromagnetic phase \cite{Men10}.

The low-energy physics of undoped antiferromagnets on a bipartite lattice is
described by an $O(3)$-symmetric non-linear $\sigma$-model \cite{Cha89}, whose
systematic treatment is realized in magnon chiral perturbation theory 
\cite{Neu89,Fis89,Has90,Has91,Has93,Chu94}. The effective theory for holes 
doped into an antiferromagnet on the square lattice was pioneered by Shraiman 
and Siggia \cite{Shr88}. In particular, these authors found an important term
in the magnon-hole action known as the Shraiman-Siggia term. Based on the
microscopic $t$-$J$ model, interesting results on magnon-mediated forces 
between holes were obtained in \cite{Kuc93} and spiral phases were studied in
\cite{Sus04,Kot04}. In analogy to baryon chiral perturbation theory for QCD 
\cite{Geo84,Gas88,Jen91,Ber92,Bec99}, a systematic low-energy effective field 
theory for magnons and holes was constructed in \cite{Kae05,Bru06}. This theory
has been used in a detailed analysis of two-hole states bound by one-magnon 
exchange \cite{Bru06,Bru06a} as well as of spiral phases \cite{Bru07}. The 
systematic effective field theory investigations have also been extended to 
electron-doped antiferromagnets \cite{Bru07a}. In that case, no 
Shraiman-Siggia-type term (with just a single spatial derivative) exists. Hence 
at low energies magnon-electron couplings are weaker than magnon-hole couplings.
As a consequence, in contrast to hole-doped systems, in electron-doped systems 
there are no spiral phases with a helical structure in the staggered 
magnetization \cite{Bru07a}.

In this paper we construct a systematic low-energy effective field theory for
hole-doped antiferromagnets on the honeycomb lattice. In the antiferromagnetic
phase, the $SU(2)_s$ spin symmetry is spontaneously broken and the fermions
pick up a mass. This is analogous to QCD where protons and neutrons pick up
their masses due to spontaneous chiral symmetry breaking. Our analysis shows
that the effective theory on the honeycomb lattice contains a term similar to
the Shraiman-Siggia term in the square lattice case \cite{Shr88}, which 
supports spiral phases. Remarkably, the leading terms of the effective theory 
have an accidental continuous rotation symmetry which is reduced to the discrete
60 degrees rotation symmetry $O$ of the microscopic honeycomb lattice only by 
the higher-order terms. While spiral phases in hole-doped antiferromagnets on 
the honeycomb lattice were explored in \cite{Jia09}, here --- as a further
application of the effective field theory method --- we derive the one-magnon 
exchange potentials between two holes and study the formation of two-hole bound 
states, which will turn out to have $f$-wave symmetry.

The rest of the paper is organized as follows. Section \ref{microscopic}
contains a symmetry analysis of the underlying Hubbard and $t$-$J$ model.
Section \ref{magnons} is devoted to the low-energy effective theory for
magnons --- in particular, a non-linear realization of the $SU(2)_s$ spin
symmetry is constructed. Based on the microscopic $t$-$J$ model, in Section
\ref{holes}, we include the holes in the effective field theory framework.
In Section \ref{potential}, one-magnon exchange potentials between two
holes are derived and the resulting two-hole bound states are investigated
in Section \ref{boundstates}. Finally, section \ref{conclusions} contains our
conclusions.

\section{Microscopic Theory}
\label{microscopic}

We assume that the Hubbard model and the $t$-$J$-model are reliable models
to describe doped quantum antiferromagnets, and therefore are valid as
concrete microscopic models for the low-energy effective field theory for
magnons and holes. Due to the fact  that the effective Lagrangian to be
constructed must inherit all symmetries of the underlying microscopic systems,
a careful symmetry analysis of these microscopic models is presented in this
section.

\subsection{Symmetries of the Honeycomb Lattice}

\begin{figure}[t]
\begin{center}
\vspace{-0.15cm}
\epsfig{file=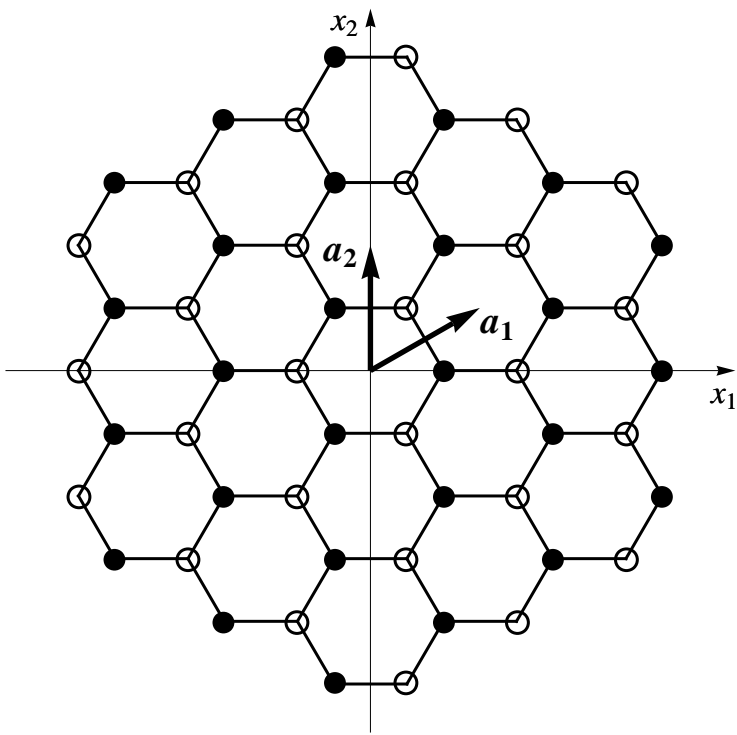,angle=0,width=10cm} \vskip-0.5cm
\end{center}
\caption{\it Bipartite non-Bravais honeycomb lattice consisting of two
triangular Bravais sublattices. The translation vectors are $a_1$ and $a_2$.}
\label{honeycomblattice}
\end{figure}

The honeycomb lattice is not a Bravais lattice --- rather, it consists of two
triangular Bravais sublattices A and B, as depicted in
Figure~\ref{honeycomblattice}. 
The primitive vectors that generate the
triangular sublattices in coordinate space are given by
\begin{equation}
a_1 = \sqrt{3} a \left(\frac{\sqrt{3}}{2},\frac{1}{2}\right), \
a_2 = \sqrt{3} a \left(0,1\right),
\end{equation}
where $a$ is the lattice spacing between two neighboring sites. The two basis
vectors $b_1$ and $b_2$ that span the reciprocal lattice obey
\begin{equation}
a_i b_j = 2 \pi \delta_{ij},
\end{equation}
and are given by
\begin{equation}
b_1 = \frac{4 \pi}{3 a} \left(1,0\right), \
b_2 = \frac{4 \pi}{3 a} \left(- \frac{1}{2},\frac{\sqrt{3}}{2}\right).
\end{equation}
The vectors $b_1$ and $b_2$ generate the hexagonal first Brillouin zone of the
triangular lattice. Since the honeycomb lattice consists of two triangular
sublattices, its momentum space is doubly-covered. 

The honeycomb lattice exhibits a number of discrete symmetries. Translations
by the vectors $a_i$ are denoted by $D_i$. Counter-clockwise rotations by 60
degrees around the center of a hexagon are denoted by $O$, and reflections at
the $x_1$-axis going through the center of the hexagon are denoted by $R$. 
Translations by other distance vectors, rotations by other angles or around 
other centers, and reflections with respect to other axes can be obtained as
combinations of the elementary symmetry operations $D_1$, $D_2$, $O$, and $R$.

\subsection{Symmetries of the Hubbard Model}
\label{hubmodel}
Let $c_{x s} ^\dagger$ denote the operator which creates a fermion with spin
$s\in\lbrace\uparrow,\downarrow\rbrace$ on a lattice site
\mbox{$x=(x_1,x_2)$}. The corresponding annihilation operator is $c_{xs}$.
These fermion operators obey the canonical anticommutation relations
\begin{equation}\label{commutators1}
\{c_{xs}\Dag,c_{ys'}\} = \delta_{xy} \delta_{ss'}, \ 
\{c_{x s},c_{y s'}\} = \{c_{x s}\Dag,c_{y s'}\Dag\} = 0.
\end{equation}
The second quantized Hubbard Hamiltonian is defined by
\begin{equation}\label{hubh}
H = - t \sum_{\ontopof{\langle x, y\rangle}{s=\uparrow ,\downarrow}}
(c_{xs} ^\dagger c_{ys} + c_{ys} ^\dagger c_{xs})
+ U\sum_x c_{x\uparrow}^\dagger c_{x\uparrow}c_{x\downarrow}^\dagger c_{x\downarrow}
- \mu ' \sum_{\ontopof{x}{s=\uparrow ,\downarrow}} c_{xs}^\dagger c_{xs},
\end{equation}
where $\langle x, y\rangle$ indicates summation over nearest neighbors, $t$
is the hopping parameter, and the parameter $U>0$ fixes the strength of the
Coulomb repulsion between two fermions located on the same lattice site. The
parameter $\mu'$ denotes the chemical potential.

The fermion creation and annihilation operators can be used to define the
following $SU(2)_s$ Pauli spinors
\begin{equation}\label{pspinor}
c_x^\dagger = \left(c_{x \uparrow}^\dagger, c_{x \downarrow}^\dagger\right), \qquad
c_x = \left(\begin{array}{c} c_{x \uparrow} \\ c_{x \downarrow} 
\end{array} \right).
\end{equation}
In terms of these operators, the Hubbard model can be reformulated as
\begin{equation}\label{hubhspin}
H = - t \sum_{\langle x y\rangle} (c_x^\dagger c_y + c_y^\dagger c_x) + 
\frac{U}{2} \sum_x (c_x^\dagger c_x - 1)^2 - \mu \sum_x (c_x^\dagger c_x - 1).
\end{equation}
The parameter $\mu = \mu' - \frac{U}{2}$ controls doping where the fermions are
counted with respect to half-filling. 

Since all terms in the effective Lagrangian must be invariant under all 
symmetries of the Hubbard model, a careful symmetry analysis of
Eq.(\ref{hubhspin}) is needed. Let us divide the symmetries of the Hubbard
model into two categories: Continuous symmetries ($SU(2)_s$, $U(1)_Q$ fermion
number and its non-Abelian extension $SU(2)_Q$), which are internal symmetries 
of Eq.(\ref{hubhspin}), and
discrete symmetries ($D_i$, $O$ and $R$), which are symmetry transformations
of the underlying honeycomb lattice. There is also time reversal which is
implemented by an anti-unitary operator $T$. This symmetry will be discussed
further in the effective field theory framework.

In order to construct the appropriate unitary transformation representing a
global $SU(2)_s$ spin rotation, we first define the total $SU(2)_s$ spin
operator by
\begin{equation}
\vec S = \sum_x \vec S_x = 
\sum_x c_x\Dag \ \frac{\vec \sigma}{2} \ c_x.
\end{equation}
The spin symmetry is implemented by the unitary operator 
\begin{equation}
V=\exp(i \vec{\eta}\cdot \vec{S}),
\end{equation}
which acts on $c_x$ as
\begin{equation}\label{trafoSU2Cspinor} 
c_x'=V\Dag c_x V=\exp(i \vec{\eta}\cdot \frac{\vec{\sigma}}{2})
c_x=gc_x,\qquad g\in SU(2)_s.
\end{equation} 
The total spin is conserved and the Hubbard Hamiltonian is invariant
under global $SU(2)_s$ spin rotations. This symmetry, however, is
spontaneously broken: the corresponding order parameter is the staggered
magnetization vector
\begin{equation}\label{stagmag}
\vec{M}_s=\sum_x (-1)^x\vec{S}_x,
\end{equation} 
which takes a non-zero expectation value in the ground state of the 
antiferromagnet. We define $(-1)^{x}=1$ for all $x \in A$ and $(-1)^{x}=-1$ for 
all  $x \in B$, where $A$ and $B$ are the two triangular sublattices of the 
honeycomb lattice.

The unitary transformation of the $U(1)_Q$ symmetry involves the charge 
operator 
\begin{equation}
Q = \sum_x Q_x = \sum_x (c_x\Dag c_x - 1)=
\sum_x (c_{x\up}\Dag c_{x\up} +c_{x\down}\Dag c_{x\down} -1),
\end{equation}
which counts the fermion number with respect to half-filling. The corresponding 
unitary operator is given by
\begin{equation}
W = \exp(i\omega Q),
\end{equation} 
and the fermion operators transform as
\begin{equation}\label{trafoU1Q}
^Qc_x = W\Dag c_x W =\exp(i\omega)c_x,\qquad \exp(i\omega)\in U(1)_Q.
\end{equation} 
Charge or fermion number are conserved due to $[H,Q] = 0$.

The Hubbard model shows invariance under shifts along the two primitive
lattice vectors $a_1$ and $a_2$. These transformations are generated by
the unitary operators $D_i$, which act on the spinor $c_x$ as
\begin{equation}\label{shiftsymm}
^{D_i}c_x = D^{\dagger}_i c_x D_i = c_{x+a_i}.
\end{equation}
By applying Eq.(\ref{shiftsymm}) on the Hubbard Hamiltonian of
Eq.(\ref{hubhspin}) and redefining the sum over lattice sites $x$, one can
see that indeed $[H,D_i] = 0$. Since the shift symmetry maps
$A \rightarrow A$ and $B \rightarrow B$, this transformation does not affect the
order parameter $\vec M_s$.

A spatial rotation by 60 degrees leaves Eq.(\ref{hubhspin}) invariant. Since
spin-orbit coupling is neglected in the Hubbard model, spin decouples from the
spatial motion and becomes an internal quantum number. The rotation symmetry
is implemented by the use of a unitary operator $O$, which acts on the
fermion operators as
\begin{equation}\label{rotationsymm}
^Oc_x = O^\dagger c_x O = c_{Ox}.
\end{equation}
Rotation symmetry on the honeycomb lattice is spontaneously broken because $O$
exchanges the two sublattices $A\leftrightarrow B$ and therefore the staggered 
magnetization
$\vec M_s$ gets flipped. This is, however, just the same as redefining the
sign of $(-1)^{x}$ and does therefore not change the physics. In the
construction of the effective field theory for magnons and holes, it will turn
out to be useful to also consider the combined symmetry $O'$ consisting of a
spatial rotation $O$ and a global $SU(2)_s$ spin rotation $g = i\sigma_2$.
$O'$ transforms $c_x$ as
\begin{equation}\label{rotationssymmp}
^{O'}c_x = O^{'\dagger} c_x O'= (i \sigma_2)\, ^Oc_x = (i \sigma_2) c_{Ox}.
\end{equation}
The specific $SU(2)_s$ element $g=i\sigma_2$ corresponds to a global spin
rotation by 180 degrees and thus flips back $\vec M_s$, such that, in fact,
at the end the order parameter is not affected by $O'$. As opposed to the
honeycomb lattice case, $\vec M_s$ changes sign under the shift symmetry $D_i$
on a bipartite square lattice \cite{Kae05}. In this case, a combined shift
symmetry $D'_i$ leaves the ground state invariant. Since on the square lattice 
a 90 degrees rotation $O$ maps sublattices $A\rightarrow A$ and 
$B\rightarrow B$, in that case the ground state is not affected by a 
rotation by an angle of 90 degrees.

Finally, the Hubbard Hamiltonian is invariant under the reflection $R$ at the 
$x_1$-axis shown in Figure 1. Under this transformation, the fermion operators 
transform as
\begin{equation}
\label{reflection}
^Rc_x = R^\dagger c_x R = c_{Rx}.
\end{equation}
Since $R$ maps the two sublattices onto themselves, $\vec M_s$ remains
invariant. 
 
In \cite{Zha90,Yan90}, Yang and Zhang proved the existence of a non-Abelian
extension of the $U(1)_Q$ fermion number symmetry in the half-filled
Hubbard model. This pseudospin symmetry contains $U(1)_Q$ as a subgroup. The
$SU(2)_Q$ symmetry is realized on the square as well as on the honeycomb 
lattice and is generated by the three operators
\begin{align}
Q^+ & = \sum_x (-1)^x c_{x \uparrow}^\dagger c_{x \downarrow}^\dagger, \qquad
Q^- = \sum_x (-1)^x c_{x \downarrow} c_{x \uparrow}, \nonumber \\
Q^3 &= \sum_x \frac{1}{2}(c_{x \uparrow}^\dagger c_{x \uparrow} +
c_{x \downarrow}^\dagger c_{x \downarrow} - 1) = \frac{1}{2} Q.
\end{align}
The factor $(-1)^x $ again distinguishes between the two sublattices $A$ and
$B$ of the honeycomb lattice. Defining $Q^1$ and $Q^2$ through
$Q^\pm = Q^1 \pm i Q^2$, one readily shows that the $SU(2)_Q$ Lie-algebra
$[Q^a,Q^b] = i \varepsilon_{abc} Q^c$, with $a, b, c \in \lbrace1,2,3\rbrace$,
indeed is satisfied  and that $[H,\vec Q]=0$ with
$\vec Q = (Q^1,Q^2,Q^3)$ for the Hubbard Hamiltonian with $\mu = 0$.

In order to write the Hubbard Hamiltonian Eq.(\ref{hubh}) or
Eq.(\ref{hubhspin}) in a manifestly invariant form under
\mbox{$SU(2)_s \times SU(2)_Q$}, we arrange the fermion operators in a
$2\times 2$ matrix-valued operator, arriving at the fermion representation
\begin{equation}\label{Coperator}
C_x = \left(\begin{array}{cc} c_{x \uparrow} &
(-1)^x \ c^\dagger_{x \downarrow} \\ c_{x \downarrow} &
- (-1)^x c_{x \uparrow}^\dagger \end{array} \right).
\end{equation}
The $SU(2)_Q$ transformation behavior of Eq.(\ref{Coperator}) can now be
worked out by applying the unitary operator
$W=\exp (i \vec \omega\cdot\vec Q)$,
\begin{equation}
^{\vec Q} C_x= W\Dag C_x W = C_x \Omega^T,
\end{equation}
with 
\begin{equation}
\Omega= \exp\left(i\vec \omega \cdot \frac{\vec \sigma}{2}\right) \in SU(2)_Q.
\end{equation}
Under an $SU(2)_s$ spin rotation, $C_x$ transforms exactly like $c_x$,
i.e.  
\begin{equation}\label{su2cx}
C_x ' = g C_x,\qquad g \in SU(2)_s.
\end{equation}
Applying an $SU(2)_s \times SU(2)_Q$ transformation to
Eq.(\ref{Coperator}) then leads to
\begin{equation}\label{trafoSU2Cspinor1} 
^{\vec Q} C_x ' = g C_x\Omega^T.
\end{equation}
Since the $SU(2)_s$ spin symmetry acts from the left and the $SU(2)_Q$
pseudospin symmetry acts from the right onto the fermion operator, it is now
obvious that these two non-Abelian symmetries commute with each other. Under 
the discrete symmetries of the Hubbard model, $C_x$ has the following 
transformation properties 
\begin{alignat}{2}
D_i:&\quad &^{D_i}C_x &= C_{x+a_i}, \nonumber \\
O:&\quad &^O C_x &= C_{Ox} \sigma_3, \nonumber \\
O':&\quad &^{O'} C_x &= \left(i\sigma_2\right) C_{Ox} \sigma_3, \nonumber \\
R:&\quad &^R C_x &= C_{Rx}.
\end{alignat}
In terms of Eq.(\ref{Coperator}), we are now able to write down the Hubbard
Hamiltonian in the manifestly $SU(2)_s$, $U(1)_Q$, $D_i$, $O$, $O'$ and
$R$ invariant form
\begin{equation}\label{HF}
H = - \frac{t}{2} \sum_{\langle x y \rangle} \mbox{Tr}[C_x^\dagger C_y +
C_y^\dagger C_x] + 
\frac{U}{12} \sum_x \mbox{Tr}[C_x^\dagger C_x C_x^\dagger C_x] -
\frac{\mu}{2} \sum_x \mbox{Tr}[C_x^\dagger C_x \sigma_3].
\end{equation}
The $\sigma_3$ Pauli matrix in the chemical potential term prevents the
Hubbard Hamiltonian from being invariant under $SU(2)_Q$ away from
half-filling. For $\mu \neq 0$, $SU(2)_Q$ is explicitly broken to its subgroup
$U(1)_Q$. In addition, the pseudospin symmetry is realized in Eq.(\ref{HF})
only for nearest-neighbor hopping. As soon as next-to-nearest-neighbor
hopping is included, the $SU(2)_Q$ invariance gets lost even for $\mu=0$. The
continuous $SU(2)_Q$ symmetry contains a discrete particle-hole symmetry. 
Although this pseudospin symmetry is not present in real materials, it will 
play an important role in the construction of the effective field theory. The 
identification of the final effective fields for holes will lead us to 
explicitly break the $SU(2)_Q$ symmetry in Section \ref{holes}.
 
\subsection{Symmetries of the $t$-$J$ Model}

Away from half-filling and for $U\gg t$, the Hubbard model reduces to the
$t$-$J$ model, which is defined by the Hamiltonian
\begin{equation}
\label{tjmodel}
H = \mathcal{P} \bigg\{- t \sum_{\langle x y\rangle} 
(c_x^\dagger c_y + c_y^\dagger c_x) +
J \sum_{\langle x y\rangle} \vec S_x \cdot \vec S_y - \mu \sum_x (c_x^\dagger
c_x - 1) \bigg\} \mathcal{P}.
\end{equation}
Using second order perturbation theory, the antiferromagnetic exchange
coupling $J$ is related to the parameters of the Hubbard model by
$J=\frac{2t^{2}}{U}>0$. Again, $t$ is the hopping amplitude, $\vec S_x$ is
the $SU(2)_s$ spin operator on a site $x$, and $\mu$ controls the doping with
 respect to a half-filled system. The projection operator $\mathcal P$ removes
all doubly occupied sites from the Hilbert space and hence the $t$-$J$ model
can only be doped with holes. In \cite{Jia08}, the single-hole sector
of the $t$-$J$ model was simulated on the honeycomb lattice by using an
efficient loop-cluster algorithm. For the construction of the effective theory
for a hole doped antiferromagnet, the $t$-$J$ model will serve as the
microscopic starting point. Except for the $SU(2)_Q$ symmetry, this model
shares all symmetries with the more general Hubbard model.

\section{Effective Theory for Magnons}

\label{magnons}

In this section we investigate the low-energy physics of an undoped quantum
antiferromagnet. We will first argue that quantum antiferromagnets are systems
featuring a spontaneous $SU(2)_s \to U(1)_s$ symmetry breakdown, which
induces two massless Goldstone bosons --- the magnons. We present the
leading-order effective action for the pure magnon sector of an
antiferromagnet on the honeycomb lattice. In addition, a  non-linear
realization of the spontaneously broken $SU(2)_s$ spin symmetry is constructed,
which will enable us to couple magnons and doped holes in section 4.

\subsection{Low-Energy Effective Action for Magnons}

In quantum antiferromagnets the symmetry group $G=SU(2)_s$ of global spin
rotations is spontaneously broken by the formation of a staggered
magnetization. The ground state of these systems is invariant only under spin
rotations in the subgroup $H=U(1)_s$. As a consequence of the spontaneous
global symmetry breaking, there are two magnons which are described by a
unit-vector field
\begin{equation}
\vec e(x) = \big(e_1(x),e_2(x),e_3(x)\big) \in S^2, \qquad \vec e(x)^2 = 1,
\end{equation}
in the coset space $G/H=SU(2)_s/U(1)_s=S^2$. Here $x=(x_1,x_2,t)$ denotes a
point in Euclidean space-time. The low-energy physics of an undoped
antiferromagnet can be completely described in terms of the field $\vec e(x)$
which represents the direction of the local staggered magnetization. 

Later, we will couple magnons to holes. Since holes have spin 1/2 and are thus
described by two-component fields, it is convenient to work with a 
$\mathbb C P(1)$ representation instead of the $O(3)$ vector
representation for the magnon field.  We introduce the $2 \times 2$ Hermitean
projection matrices $P(x)$ defined by
\begin{equation}\label{Pdefinition}
P(x) = \frac{1}{2}\big[\mathbbm{1} + \vec e(x) \cdot \vec \sigma\big]
=\frac{1}{2}
\left(\begin{array}{cc} 
1+e_3(x) & e_1(x)-ie_2(x)\\ 
e_1(x)+ie_2(x) & 1-e_3(x) 
\end{array} \right),
\end{equation}
obeying
\begin{equation}
P(x)^{\dagger} = P(x), \qquad  \mbox{Tr} P(x) = 1, \qquad P(x)^2 = P(x).
\end{equation} 
In terms of $P(x)$, to lowest-order in a systematic derivative expansion, the
effective action for magnons is given by
\begin{equation}\label{actionP} 
S[P] = \int d^2\!x \, dt \ \rho_s \mbox{Tr} [ \p_i P \p_i P +
\frac{1}{c^2} \p_t P \p_t P ].
\end{equation}
Here we have introduced two low-energy constants, the spin stiffness $\rho_s$ 
and the spinwave velocity $c$. The values of these low-energy constants have 
been determined very precisely using Monte Carlo
simulations \cite{Wie94,Bea96,Ger09}. It should be pointed out that this
leading-order contribution to the effective action is exactly the same as for
an antiferromagnet on a square lattice. Deviations will only show up when
higher order terms with more derivatives are considered.

We now discuss how the magnon field $P(x)$ transforms under the various
symmetries of the underlying microscopic models. Under global $SU(2)_s$ spin
transformations the staggered magnetization field transforms as
\begin{equation}
P(x)' = g P(x) g^\dagger.
\end{equation} 
Note that it is invariant under the Abelian and the non-Abelian fermion number
symmetries $U(1)_Q$ and $SU(2)_Q$, i.e.
\begin{equation}
^{\vec Q} P(x) = P(x).
\end{equation}
Under the displacement $D_i$ and the reflection symmetry $R$, the sublattices 
are not interchanged such that
\begin{align}
^{D_i} P(x)&= P(x), \nonumber \\
^R P(x)&= P(Rx).
\end{align}
Under a rotation by 60 degrees, the staggered magnetization vector changes
sign, i.e.\ $^O \vec e(x) = - \vec e(Ox)$, and therefore
\begin{equation}\label{OP}
^O P(x) = \frac{1}{2} \big[\mathbbm{1} - \vec e(Ox) \cdot \vec \sigma\big] =
\mathbbm{1} - P(Ox).
\end{equation} 
Note that in an antiferromagnet on the honeycomb lattice the 60 degrees
rotation symmetry is spontaneously broken, whereas in an antiferromagnet on
the square lattice, it is the displacement symmetry by one lattice spacing
which is spontaneously broken. The above transformation property simplifies
under the composed symmetry $O'$,
\begin{equation}\label{OpP}
^{O'} P(x) = (i \sigma_2) \, ^O P(x) (i \sigma_2)^\dagger = P(Ox)^*.
\end{equation}
Under time-reversal $T$, which turns a space-time point $x = (x_1,x_2,t)$ into
$Tx = (x_1,x_2,-t)$, the staggered magnetization changes sign and, as a
consequence,
\begin{equation}
^T P(x)=\mathbbm{1}-P(Tx).
\end{equation}
Since also $T$ is a spontaneously broken symmetry, again it is useful to 
consider the
composed transformation $T'$ consisting of a regular time-reversal $T$ and the
specific spin rotation $g = i \sigma_2$. Under the unbroken symmetry $T'$ the
magnon field $P(x)$ transforms as
\begin{equation}
^{T'} P(x) = (i \sigma_2)^T P(x) (i \sigma_2)^\dagger = P(Tx)^*.
\end{equation}
The effective action in Eq.(\ref{actionP}) is invariant under all these
symmetries.

\subsection{Non-linear Realization of the $SU(2)_s$ Symmetry}

In order to couple the fermions to the magnons, i.e.\ to the antiferromagnetic
order parameter, a non-linear realization of the $SU(2)_s$ symmetry has been
constructed and discussed in detail in \cite{Kae05}. The spin symmetry is
implemented on the fermion fields by a non-linear local transformation
$h(x) \in U(1)_s$. This local transformation is constructed from the global
transformation $g \in SU(2)_s$ and the magnon field $P(x)$ as follows. One
first defines a local, unitary transformation $u(x) \in SU(2)_s$ which
diagonalizes the staggered magnetization field, i.e.
\begin{equation}
\label{udiag}
u(x) P(x) u(x)^\dagger = \frac{1}{2}(\1 + \sigma_3) = 
\left(\begin{array}{cc} 
1 & 0 \\ 
0 & 0 
\end{array} \right),
\qquad u_{11}(x)\geq 0.
\end{equation}
In order to make $u(x)$ uniquely defined, we demand that the element 
$u_{11}(x)$ is real and non-negative. Using Eq.(\ref{Pdefinition}) and
spherical coordinates for $\vec e(x)$, i.e.
\begin{equation}
\label{epolar}
\vec e(x) = \big(\sin \theta(x) \cos \varphi(x), \sin \theta(x) \sin
\varphi(x), \cos \theta(x)\big), 
\end{equation}
one obtains \cite{Kae05}
\begin{align}
\label{umatrixdef}
u(x)&= \frac{1}{\sqrt{2(1+e_3(x))}}
\left(\begin{array}{cc} 1 + e_3(x) & e_1(x) - i e_2(x) \\
- e_1(x) - i e_2(x) & 1 + e_3(x) \end{array} \right) \nonumber \\[1ex]
&= \left(\begin{array}{cc}
\cos\left(\frac{\theta(x)}{2}\right) & \sin\left(\frac{\theta(x)}{2}\right)
\exp(-i\varphi(x))
\\[0.5ex]
-\sin\left(\frac{\theta(x)}{2}\right)\exp (i\varphi(x)) & 
\cos\left(\frac{\theta(x)}{2}\right)
 \end{array} \right).
\end{align}
Note that the local transformation $u(x)$ rotates an arbitrary staggered
magnetization field configuration $P(x)$ into the specific
constant diagonal field configuration with $P(x)=\frac{1}{2}(\1+\sigma_3)$.
Under a global $SU(2)_s$ transformation $g$ the diagonalizing field $u(x)$
transforms as
\begin{equation}\label{trafoumatrix} 
u(x)'=h(x)u(x)g^{\dagger}, \qquad u_{11}(x)' \geq 0,
\end{equation} 
which implicitly defines the non-linear symmetry transformation
\begin{equation}\label{defh}
h(x)=\exp\big(i \alpha(x)\sigma_3\big)= 
\left( \begin{array}{cc}
\exp(i \alpha(x)) & 0 \\
0 & \exp (-i \alpha(x))
\end{array} \right)
\in U(1)_s.
\end{equation}
The transformation $h(x)$ is uniquely defined since we demand that
$u_{11}(x)'$ is again real and non-negative. 

The transformation behavior of the field $u(x)$ can be easily worked out from
the known transformation behavior of $P(x)$. Since $u(x)$ contains only magnon
degrees of freedom, it transforms trivially under both the Abelian and the
non-Abelian fermion number symmetries $U(1)_Q$ and $SU(2)_Q$, i.e.
\begin{equation}
^{\vec Q} u(x)= u(x).
\end{equation}
Under the displacement $D_i$ and the reflection symmetry $R$ one finds
\begin{align}
^{D_i}u(x)&=u(x), \nonumber \\
^Ru(x)&=u(Rx).
\end{align} 
The spontaneous breaking of the 60 degrees rotation symmetry $O$ which takes
$\vec{e}(x)$ to $-\vec{e}(Ox)$ leads to
\begin{equation}
^O u(x)=\tau(Ox)u(Ox),
\end{equation}
with
\begin{align}
\label{taueq}
\tau(x)&= \frac{1}{\sqrt{e_1(x)^2+e_2(x)^2}}
\left( 
\begin{array}{cc}
0 & -e_1(x) + ie_2(x) \\
e_1(x) +ie_2(x) & 0
\end{array} 
\right)
\nonumber \\[1ex]
&=
\left( 
\begin{array}{cc}
0 & -\exp(-i \varphi(x))\\
\exp(i\varphi(x)) & 0
\end{array} 
\right).
\end{align}
Under the combined symmetry $O'$ one finds
\begin{equation}\label{Oprimeu}
^{O'}u(x)=u(Ox)^{*}.
\end{equation}
Since time-reversal $T$ is a spontaneously broken discrete symmetry in an
antiferromagnet, it acts on $u(x)$ as
\begin{equation}
^{T}u(x)=
\tau(Tx)u(Tx).
\end{equation}
On the other hand, the combined time-reversal $T'$ is unbroken and therefore
realized in a linear manner, i.e.
\begin{equation}
^{T'}u(x)=
u(Tx)^{*}.
\end{equation}

Finally, we introduce the composite magnon fields $v_\mu(x)$ whose components
will be used to couple the magnons to the fermions. Using the diagonalizing 
field $u(x)$, we define the composite magnon field
\begin{equation}\label{defvmu}
v_\mu(x)= u(x) \p_\mu u(x)^{\dagger},
\end{equation}
which under $SU(2)_s$ transforms as
\begin{equation}\label{trafov}
v_\mu(x)' = h(x) u(x) g^{\dagger} \p_\mu [g u(x)^{\dagger} h(x)^{\dagger} ] =
h(x) [v_\mu(x) + \p_\mu] h(x)^{\dagger}.
\end{equation}
Since the field $v_\mu(x)$ is traceless and anti-Hermitean, it can be written
as a linear combination of the Pauli matrices $\sigma_a$,
\begin{equation}
v_{\mu}(x)=iv_{\mu}^a(x) \sigma_a, \qquad a\in \{1,2,3\}, \qquad v_\mu^a(x) \in
\mathbbm{R}.
\end{equation}
Introducing
\begin{equation}
v_{\mu}^\pm(x)=v_{\mu}^1(x)\mp i v_{\mu}^2(x),
\end{equation}
we arrive at
\begin{equation}\label{vmumatrix}
v_{\mu}(x)=
i
\left( \begin{array}{cc}
v_\mu^3(x) & v_\mu^+(x) \\
v_\mu^-(x) & -v_\mu^3(x)
\end{array}\right).
\end{equation}
Under global $SU(2)_s$ transformations the components of $v_{\mu}$ transform as
\begin{eqnarray}\label{gaugevmu}
v_{\mu}^3(x)'&=&v_{\mu}^3(x)-\p_{\mu}\alpha(x), \nonumber\\
v_{\mu}^\pm(x)'&=&\exp(\pm 2i\alpha(x))  v_{\mu}^\pm(x),
\end{eqnarray}
which indicates that $v_{\mu}^3$ behaves like an Abelian $U(1)_s$ gauge field,
while $v^\pm_\mu(x)$ exhibit the behavior of vector fields ``charged'' under
$U(1)_s$. The transformation properties of the components $v_\mu^3(x)$ and
$v^\pm_\mu(x)$ under the discrete symmetries can be worked out from the
definition of $v_\mu(x)$ in Eq.(\ref{defvmu}) as well, and are summarized as
follows
\begin{alignat}{2}\label{trafovmu3}
D_i:&\quad &^{D_i}v_{\mu}^{3}(x) &= v_{\mu}^{3}(x), \nonumber\\
O:&\quad &^Ov_{1}^{3}(x) &= \tfrac{1}{2}\big[-v_{1}^{3}(Ox)+\partial_{1}
\varphi(Ox)-\sqrt{3}v_{2}^{3}(Ox)+\sqrt{3}\partial_{2}\varphi(Ox)\big],
\nonumber \\
&\quad &^Ov_{2}^{3}(x) &= \tfrac{1}{2}\big[\sqrt{3}v_{1}^{3}(Ox)-\sqrt{3}
\partial_{1}\varphi(Ox)-v_{2}^{3}(Ox)+\partial_{2}\varphi(Ox)\big], \nonumber \\
&\quad &^Ov_{t}^{3}(x) &=-v_{t}^{3}(Ox)+\partial_{t}\varphi(Ox), \nonumber \\
O':&\quad &^{O'}v_{1}^{3}(x) &= -\tfrac{1}{2}\big[v_{1}^{3}(Ox)+
\sqrt{3}v_{2}^{3}(Ox)\big], \nonumber \\
&\quad &^{O'}v_{2}^{3}(x) &= \tfrac{1}{2}\big[\sqrt{3}v_{1}^{3}(Ox)-v_{2}^{3}(Ox)
\big], \nonumber \\
&\quad &^{O'}v_{t}^{3}(x) &= -v_{t}^{3}(Ox), \nonumber \\
R&:\quad &^Rv_{1}^{3}(x) &=v_{1}^{3}(Rx), \nonumber \\
&\quad &^Rv_{2}^{3}(x) &= -v_{2}^{3}(Rx), \nonumber \\
&\quad &^Rv_{t}^{3}(x) &= v_{t}^{3}(Rx), \nonumber \\
T&:\quad &^Tv_{i}^{3}(x) &=-v_{i}^{3}(Tx)+\partial_{i}\varphi(Tx), \nonumber \\
&\quad &^Tv_{t}^{3}(x) &=v_{t}^{3}(Tx)-\partial_{t}\varphi(Tx), \nonumber \\
T'&:\quad &^{T'}v_{i}^{3}(x) &=-v_{i}^{3}(Tx), \nonumber \\
&\quad &^{T'}v_{t}^{3}(x) &= v_{t}^{3}(Tx),
\end{alignat}
and
\begin{alignat}{2}\label{trafovmupm}
D_i:&\quad &^{D_i}v_{\mu}^{\pm}(x) &= v_{\mu}^{\pm}(x), \nonumber \\
O:&\quad &^Ov_{1}^{\pm}(x) &= -\exp(\mp 2i\varphi(Ox))\tfrac{1}{2}
\big(v_{1}^{\mp}(Ox)+\sqrt{3}v_{2}^{\mp}(Ox)\big), \nonumber \\
&\quad &^Ov_{2}^{\pm}(x) &= \exp(\mp 2i\varphi(Ox))\tfrac{1}{2}
\big(\sqrt{3}v_{1}^{\mp}(Ox)-v_{2}^{\mp}(Ox)\big), \nonumber \\
&\quad &^Ov_{t}^{\pm}(x) &=-\exp(\mp 2i\varphi(Ox))v_{t}^{\mp}(x), \nonumber \\
O':&\quad &^{O'}v_{1}^{\pm}(x) &= -\tfrac{1}{2}\big(v_{1}^{\mp}(Ox)
+\sqrt{3}v_{2}^{\mp}(Ox)\big), \nonumber \\
&\quad &^{O'}v_{2}^{\pm}(x) &= \tfrac{1}{2}
\big(\sqrt{3}v_{1}^{\mp}(Ox)-v_{2}^{\mp}(Ox)\big), \nonumber \\
&\quad &^{O'}v_{t}^{\pm}(x) &= -v_{t}^{\mp}(Ox), \nonumber \\
R&:\quad &^Rv_{1}^{\pm}(x) &=v_{1}^{\pm}(Rx), \nonumber \\
&\quad &^Rv_{2}^{\pm}(x) &= -v_{2}^{\pm}(Rx), \nonumber \\
&\quad &^Rv_{t}^{\pm}(x) &= v_{t}^{\pm}(Rx), \nonumber \\
T&:\quad &^Tv_{i}^{\pm}(x) &=-\exp(\mp 2i\varphi(Tx))v_{i}^{\mp}(Tx), 
\nonumber \\
&\quad &^Tv_{t}^{\pm}(x) &=\exp(\mp 2i\varphi(Tx))v_{t}^{\mp}(Tx), \nonumber \\
T'&:\quad &^{T'}v_{i}^{\pm}(x) &=-v_{i}^{\mp}(Tx), \nonumber \\
&\quad &^{T'}v_{t}^{\pm}(x) &= v_{t}^{\mp}(Tx).
\end{alignat}
The magnon action of Eq.(\ref{actionP}) can now be reformulated in terms
of the composite magnon field $v_{\mu}(x)$,
\begin{equation}
\label{vmagnonaction}
S[v_\mu^\pm] = \int d^2\!x \, dt \ 2\rho_s \left(v^+_i v^-_i + \frac{1}{c^2}
v_t^+ v_t^-\right).
\end{equation}
At a first glance, the expression $v_{\mu}^{+}v_{\mu}^{-}$ looks like a mass 
term of a charged vector field. However, since it contains derivatives acting 
on $u(x)$, it is just the kinetic term of a massless Goldstone boson.

\section{Effective Theory for Magnons and Holes}
\label{holes}
In this section we construct a systematic low-energy effective theory for
holes coupled to magnons. As a first step toward building the effective
theory, we identify the correct low-energy degrees of freedom that describe
the holes. Then the transformation behavior of these fermionic fields is
investigated in great detail. Finally, the most general effective Lagrangian
for magnons and holes is constructed.

\subsection{Fermion Fields and their Transformation Properties}

\begin{figure}[t]
\begin{center}
\vspace{-0.15cm}
\epsfig{file=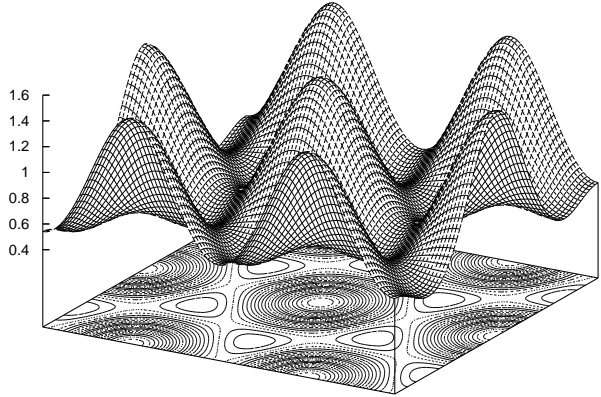,angle=-90,width=10cm} \vskip-0.5cm
\end{center}
\caption{\it The dispersion relation $E(k)/t$ for a single hole in an
antiferromagnet on the honeycomb lattice simulated in the $t$-$J$ model for
$J/t=2$ \cite{Jia08}.}
\label{single_landscape}
\end{figure}

\begin{figure}[t]
\begin{center}
\epsfig{file=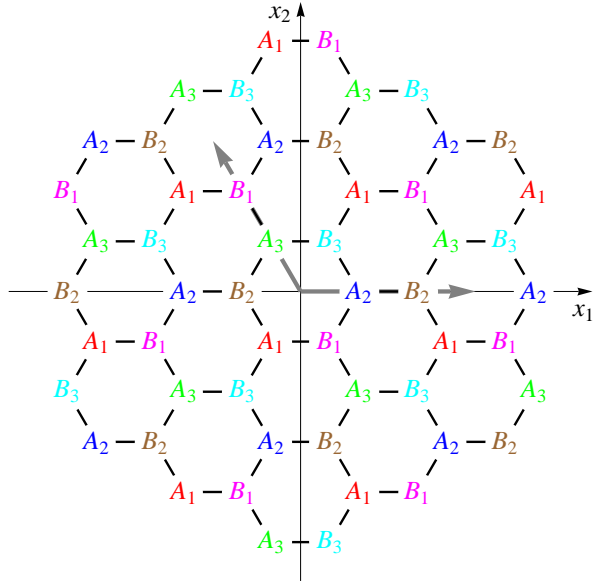,width=8cm} 
\end{center}
\caption{\it $\lbrace A_{1}, A_{2}, A_{3}\rbrace$ and $\lbrace B_{1}, B_{2},
B_{3}\rbrace$ sublattice structure and the corresponding primitive lattice
vectors.}
\label{sublattice}
\end{figure}

In order to construct the effective theory for hole-doped antiferromagnets, it
is essential to know where the hole pockets are located in momentum space.
The dispersion relation $E(k)$ for a single hole in the $t$-$J$ model on the
honeycomb lattice was simulated using an efficient loop-cluster
algorithm \cite{Jia08}. The result is shown in Figure~\ref{single_landscape}.
This simulation clearly shows spherically shaped hole pockets centered around
\mbox{$(\pm\frac{2\pi}{3a},\pm\frac{2\pi}{3\sqrt{3} a})$} and
\mbox{$(0,\pm \frac{4\pi}{3\sqrt{3} a})$} in the first Brillouin zone.
Therefore, doped holes occupy the two pockets $\alpha$ and $\beta$ with
lattice momenta
\begin{equation}
k^{\alpha}=-k^{\beta}=(0,\frac{4\pi}{3\sqrt{3}a}).
\end{equation}
Together with the origin, these two points form a minimal set of three points
in momentum space. The three points in coordinate space that are related to
$0,k^\alpha,k^\beta$ by a discrete Fourier transform, define three triangular
sublattices $A_1$, $A_2$, and $A_3$, as well as $B_1$, $B_2$, and $B_3$ on the
$A$- and $B$-sublattices of the honeycomb lattice. The geometry of these six
triangular sublattices is illustrated in Figure~\ref{sublattice}.
We now introduce fermionic lattice operators with a sublattice index $X$ as an
intermediate step between the microscopic and the effective fermion fields,
\begin{equation}\label{Bruecke}
\Psi^{X}_x = u(x) C_x,
\end{equation}
with $x\in X$, $X \in \lbrace A_{1}, A_{2}, A_{3}, B_{1}, B_{2}, B_{3}\rbrace$.
The above definition of $\Psi^{X}_x$ contains the diagonalizing matrix $u(x)$
of Eq.(\ref{umatrixdef}) and hence accounts for the non-linearly realized
$SU(2)_s$ symmetry on the effective fermion fields. On even and odd
sublattices the fermion operator has the following components
\begin{equation}\label{BrueckeA}
\Psi^X_x=u(x)
\begin{pmatrix}
c_{x\up} & c_{x\down}\Dag\\
c_{x\down} & -c_{x\up}\Dag
\end{pmatrix}
=
\begin{pmatrix}
\psi^X_{x,+}& \psi^{X\Dag}_{x,-}\\
\psi^X_{x,-}& -\psi^{X\Dag}_{x,+}
\end{pmatrix},
\quad  x\in X, X \in \lbrace A_{1}, A_{2}, A_{3}\rbrace ,
\end{equation}
and
\begin{equation}\label{BrueckeB}
\Psi^X_x=u(x)
\begin{pmatrix}
c_{x\up} & -c_{x\down}\Dag\\
c_{x\down} & c_{x\up}\Dag
\end{pmatrix}
=
\begin{pmatrix}
\psi^X_{x,+}& -\psi^{X\Dag}_{x,-}\\
\psi^X_{x,-}& \psi^{X\Dag}_{x,+}
\end{pmatrix},
\quad  x\in X, X \in \lbrace B_{1}, B_{2}, B_{3}\rbrace.
\end{equation}
Note that with the spontaneously broken spin symmetry only the spin direction
relative to the local staggered magnetization is still a good quantum number.
The subscript $+(-)$ then indicates anti-parallel (parallel) spin alignment
with respect to the direction of $\vec e(x)$. According to
Eqs.~(\ref{su2cx}) and (\ref{trafoumatrix}), under the $SU(2)_s$ symmetry
one obtains
\begin{equation}\label{spintrafoPsisublattice}
{\Psi_x^X}' = u(x)' C_x' = h(x) u(x) g^\dagger g C_x = h(x) \Psi_x^X .
\end{equation}
Similarly, under the $SU(2)_{Q}$ symmetry one finds
\begin{equation}
^{\vec Q}\Psi_x^X = {}^{\vec Q}u(x) ^{\vec Q}C_x = u(x) C_x \Omega^T 
= \Psi_x^X \Omega^T. 
\end{equation}
The discrete symmetries are implemented on the above fermionic lattice
operators $\Psi_x^{X}$ as
\begin{alignat}{2}\label{trafofermioniclattice}
D_i:&\quad &^{D_i}\Psi_x^X &=\Psi_{x+a_{i}}^{D_{i}X}, \nonumber \\
O:&\quad &^O\Psi_x^X &= \tau(Ox)\Psi_{Ox}^{OX}\sigma_{3}, \nonumber \\
O':&\quad &^{O'}\Psi_x^X &= (i\sigma_{2})\Psi_{Ox}^{OX}\sigma_{3}, \nonumber \\
R:&\quad &^R\Psi_x^X &= \Psi_{Rx}^{RX}. 
\end{alignat}

In the effective theory doped holes are described by anticommuting
matrix-valued Grassmann fields
\begin{align}\label{PsiGrass}
&\Psi^X(x) = \left(\begin{array}{cc} \psi^X_+(x) & \psi^{X\dagger}_-(x) \\ 
\psi^X_-(x) & - \psi^{X\dagger}_+(x) \end{array} \right), \qquad
  X \in\lbrace A_{1},A_{2},A_{3}\rbrace, \nonumber \\
&\Psi^X(x) = \left(\begin{array}{cc} \psi^X_+(x) & 
- \psi^{X\dagger}_-(x) \\ 
\psi^X_-(x) & \psi^{X\dagger}_+(x) \end{array} \right), \qquad
  X \in\lbrace B_{1},B_{2},B_{3}\rbrace ,
\end{align}
consisting of Grassmann field components $\psi_{\pm}^{X}(x)$ instead of
lattice operators $\psi_{x,\pm}^{X}$. We also introduce
\begin{eqnarray}
\label{phidagger}
&&\Psi^{X\dagger}(x) = \left(\begin{array}{cc} 
\psi^{X\dagger}_+(x) & \psi^{X\dagger}_-(x) \\ \psi^X_-(x) & - \psi^X_+(x) 
\end{array} \right), \qquad X \in\lbrace A_{1},A_{2},A_{3}\rbrace, \nonumber \\
&&\Psi^{X\dagger}(x) = \left(\begin{array}{cc}
\psi^{X\dagger}_+(x) & \psi^{X\dagger}_-(x) \\ - \psi^X_-(x) & \psi^X_+(x)
\end{array} \right), \qquad X \in\lbrace B_{1},B_{2},B_{3}\rbrace ,
\end{eqnarray}
consisting of the same Grassmann fields as $\Psi^{X}(x)$. Therefore,
$\Psi^{X\dagger}(x)$ is not independent of $\Psi^{X}(x)$. By postulating that
the matrix-valued fields $\Psi^X(x)$ transform exactly as the lattice
operator $\Psi^X_x$, one obtains
\begin{alignat}{2}\label{trafofermionicmatrix}
SU(2)_{s}:&\quad &\Psi^{X}(x)' &= h(x)\Psi^{X}(x), \nonumber\\
SU(2)_{Q}:&\quad &^{\vec Q}\Psi^{X}(x)&=\Psi^{X}(x)\Omega^{T}, \nonumber\\
D_i:&\quad &^{D_i}\Psi^{X}(x) &=\Psi^{D_{i}X}(x), \nonumber \\
O:&\quad &^O\Psi^{X}(x) &= \tau(Ox)\Psi^{OX}(Ox)\sigma_{3}, \nonumber \\
O':&\quad &^{O'}\Psi^{X}(x) &= (i\sigma_{2})\Psi^{OX}(Ox)\sigma_{3}, \nonumber \\
R:&\quad &^R\Psi^{X}(x) &= \Psi^{RX}(Rx), \nonumber\\
T:&\quad &^T\Psi^X(x) &= 
\tau(Tx) (i \sigma_2) \left[\Psi^{X\dagger}(Tx)^T\right] \sigma_3, \nonumber \\
  &\quad &^T\Psi^{X\dagger}(x) &=
 - \sigma_3 \left[\Psi^X(Tx)^T\right] (i \sigma_2)^\dagger \tau(Tx)^\dagger, 
\nonumber \\
T':&\quad &^{T'}\Psi^X(x) &= - \left[\Psi^{X\dagger}(Tx)^T\right] \sigma_3, 
\nonumber \\
   &\quad &^{T'}\Psi^{X\dagger}(x) &= \sigma_3 \left[\Psi^X(Tx)^T\right].
\end{alignat}
Here the transformation behavior under time-reversal $T$ and $T'$ is also 
listed.
The form of the time-reversal symmetry $T$ for an effective field theory with
a non-linearly realized $SU(2)_{s}$ symmetry can be deduced from the canonical
form of time-reversal in the path integral of a non-relativistic theory with a
linearly realized spin symmetry. The fermion fields in the two formulations
just differ by a factor $u(x)$. Note, that an upper index $T$ on the left
denotes time-reversal, while on the right it denotes transpose. In components
the transformation rules take the form
\begin{alignat}{2}\label{trafofermioniccomponents}
SU(2)_s:&\quad &\psi^X_\pm(x)' &= \exp(\pm i \alpha(x)) \psi^X_\pm(x),
\nonumber \\
U(1)_Q:&\quad &^Q\psi^X_\pm(x) &= \exp(i \omega) \psi^X_\pm(x),
\nonumber \\
D_i:&\quad &^{D_i}\psi^X_\pm(x) &= \psi^{D_iX}_\pm(x), \nonumber \\
O:&\quad &^{O}\psi^X_\pm(x) &=\mp \exp(\mp i\varphi(Ox))\psi^{OX}_\mp(Ox),
\nonumber \\
O':&\quad &^{O'}\psi^X_\pm(x) &=\pm \psi^{OX}_\mp(Ox), \nonumber \\
R:&\quad &^R\psi^X_\pm(x) &= \psi^{RX}_\pm(Rx), \nonumber \\
T:&\quad &^T\psi^X_\pm(x) &= \exp(\mp i \varphi(Tx)) \psi^{X\dagger}_\pm(Tx),
\nonumber \\
&\quad &^T\psi^{X\dagger}_\pm(x) &= - \exp(\pm i \varphi(Tx)) \psi^X_\pm(Tx),
\nonumber \\
T':&\quad &^{T'}\psi^X_\pm(x) &= - \psi^{X\dagger}_\pm(Tx), \nonumber \\
&\quad &^{T'}\psi^{X\dagger}_\pm(x) &= \psi^X_\pm(Tx).
\end{alignat}
Since the spin as well as the staggered magnetization get flipped under
time-reversal, the projection of one onto the other remains invariant.

\begin{figure}[htbp!]
\begin{center}
\epsfig{file=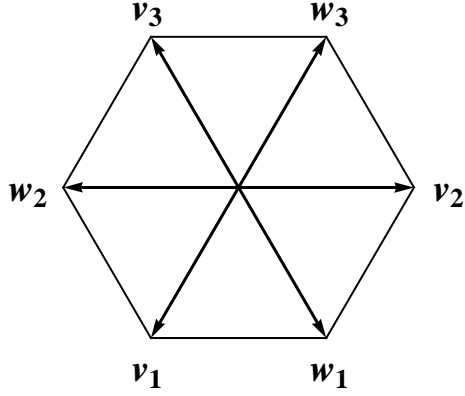,width=6cm} 
\end{center}
\caption{\it Sublattice vectors from Eq.(\ref{sublatticevectors}).}
\label{vectors}
\end{figure}

We now want to directly relate the fermion fields to the lattice momenta
$k^{\alpha}$ and $k^{\beta}$, i.e.~to the hole pockets $\alpha$ and $\beta$. The
new degrees of freedom are thus labeled with an additional ``flavor'' index
$f\in\lbrace \alpha,\beta\rbrace$. These fields are defined using the
following discrete Fourier transformations 
\begin{eqnarray}
\label{discft}
\psi^{A,f}(x)&=&\frac{1}{\sqrt{3}}\sum_{n=1}^{3} \exp(-ik^{f}v_n)\psi^{A_{n}}(x),
\nonumber \\
\psi^{B,f}(x)&=&\frac{1}{\sqrt{3}}\sum_{n=1}^{3} \exp(-ik^{f}w_n)\psi^{B_{n}}(x),
\end{eqnarray}
where 
\begin{align}\label{sublatticevectors}
v_1 &= \left( -\tfrac{1}{2}a , -\tfrac{\sqrt{3}}{2}a \right),
& v_2 &= \left( a , 0 \right),
& v_3 &= \left( -\tfrac{1}{2}a , \tfrac{\sqrt{3}}{2}a \right), \nonumber \\
w_1 &= \left( \tfrac{1}{2}a , -\tfrac{\sqrt{3}}{2}a \right),
& w_2 &= \left( -a , 0 \right),
& w_3&= \left( \tfrac{1}{2}a , \tfrac{\sqrt{3}}{2}a \right).
\end{align}
The above vectors connect the discrete three-sublattice structure of $A$ and
$B$ in position space with lattice momenta $k^{f}$ in momentum space
(Figure~\ref{vectors}). 
The fields with the pocket (or momentum) index then read
\begin{eqnarray}\label{lincommatrix}
\Psi^{A,\alpha}(x)&=&\frac{1}{\sqrt{3}}\Big[\exp\left(i\tfrac{2\pi}{3}\right)
\Psi^{A_{1}}(x)+\Psi^{A_{2}}(x)+
\exp\left(-i\tfrac{2\pi}{3}\right)\Psi^{A_{3}}(x)\Big], \nonumber\\
\Psi^{A,\beta}(x)&=&\frac{1}{\sqrt{3}}\Big[\exp\left(-i\tfrac{2\pi}{3}\right)
\Psi^{A_{1}}(x)+\Psi^{A_{2}}(x)+\exp\left(i\tfrac{2\pi}{3}\right)
\Psi^{A_{3}}(x)\Big], \nonumber\\
\Psi^{B,\alpha}(x)&=&\frac{1}{\sqrt{3}}\Big[\exp\left(i\tfrac{2\pi}{3}\right)
\Psi^{B_{1}}(x)+\Psi^{B_{2}}(x)+
\exp\left(-i\tfrac{2\pi}{3}\right)\Psi^{B_{3}}(x)\Big], \nonumber\\
\Psi^{B,\beta}(x)&=&\frac{1}{\sqrt{3}}\Big[\exp\left(-i\tfrac{2\pi}{3}\right)
\Psi^{B_{1}}(x)+\Psi^{B_{2}}(x)+
\exp\left(i\tfrac{2\pi}{3}\right)\Psi^{B_{3}}(x)\Big].
\end{eqnarray}
The Fourier transformed matrix-valued fields of
Eq.(\ref{lincommatrix}) can be written as
\begin{align}\label{PsiGrassft}
&\Psi^{A,f}(x) =\frac{1}{\sqrt{3}}\sum_{n=1}^{3} \exp(-ik^{f}v_n)\Psi^{A_{n}}(x)=
\left(\begin{array}{cc} \psi^{A,f}_{+}(x) & \psi^{A,f'\dagger}_{-}(x) \\ 
\psi^{A,f}_{-}(x) & - \psi^{A,f'\dagger}_{+}(x) \end{array} \right), \nonumber \\
&\Psi^{B,f}(x) =\frac{1}{\sqrt{3}}\sum_{n=1}^{3} \exp(-ik^{f}w_n)\Psi^{B_{n}}(x)=
\left(\begin{array}{cc} \psi^{B,f}_{+}(x) & -\psi^{B,f'\dagger}_{-}(x) \\ 
\psi^{B,f}_{-}(x) &  \psi^{B,f'\dagger}_{+}(x) \end{array} \right),
\end{align}
with their conjugated counterparts
\begin{align}\label{PsiGrassftdagger}
&\Psi^{A,f\dagger}(x) =\left(\begin{array}{cc} \psi^{A,f\dagger}_{+}(x) & 
\psi^{A,f\dagger}_{-}(x) \\ 
\psi^{A,f'}_{-}(x) & - \psi^{A,f'}_{+}(x) \end{array} \right),
\quad
&\Psi^{B,f\dagger}(x) =\left(\begin{array}{cc} \psi^{B,f\dagger}_{+}(x) & 
\psi^{B,f\dagger}_{-}(x) \\ 
-\psi^{B,f'}_{-}(x) &  \psi^{B,f'}_{+}(x) \end{array} \right).
\end{align}
The transformation properties of the fields in Eq.(\ref{lincommatrix}) are
\begin{alignat}{2}\label{trafofermionicmatrixsubmom}
SU(2)_{s}:&\quad &\Psi^{X,f}(x)' &= h(x)\Psi^{X,f}(x), \nonumber\\
SU(2)_{Q}:&\quad &^{\vec Q}\Psi^{X,f}(x)&=\Psi^{X,f}(x)\Omega^{T}, \nonumber\\
D_i:&\quad &^{D_i}\Psi^{X,f}(x) &=\exp(ik^{f}a_{i})\Psi^{X,f}(x), \nonumber \\
O:&\quad &^O\Psi^{A,\alpha}(x) &= \exp(-i\tfrac{2\pi}{3})\tau(Ox)\Psi^{B,\beta}
(Ox)\sigma_{3}, \nonumber \\
&\quad &^O\Psi^{A,\beta}(x)&= \exp(i\tfrac{2\pi}{3})\tau(Ox)\Psi^{B,\alpha}
(Ox)\sigma_{3}, \nonumber \\
&\quad &^O\Psi^{B,\alpha}(x) &= \exp(i\tfrac{2\pi}{3})\tau(Ox)\Psi^{A,\beta}
(Ox)\sigma_{3}, \nonumber \\
&\quad &^O\Psi^{B,\beta}(x) &= \exp(-i\tfrac{2\pi}{3})\tau(Ox)\Psi^{A,\alpha}
(Ox)\sigma_{3}, \nonumber \\
O':&\quad &^{O'}\Psi^{A,\alpha}(x) &= \exp(-i\tfrac{2\pi}{3})(i\sigma_{2})
\Psi^{B,\beta}(Ox)\sigma_{3}, \nonumber \\
&\quad &^{O'}\Psi^{A,\beta}(x)&= \exp(i\tfrac{2\pi}{3})(i\sigma_{2})
\Psi^{B,\alpha}(Ox)\sigma_{3}, \nonumber \\
&\quad &^{O'}\Psi^{B,\alpha}(x) &= \exp(i\tfrac{2\pi}{3})(i\sigma_{2})
\Psi^{A,\beta}(Ox)\sigma_{3}, \nonumber \\
&\quad &^{O'}\Psi^{B,\beta}(x) &= \exp(-i\tfrac{2\pi}{3})(i\sigma_{2})
\Psi^{A,\alpha}(Ox)\sigma_{3}, \nonumber \\
R:&\quad &^R\Psi^{X,f}(x) &= \Psi^{X,f'}(Rx), \nonumber\\
T:&\quad &^T\Psi^{X,f}(x) &= 
\tau(Tx) (i \sigma_2) \left[\Psi^{X,f'\dagger}(Tx)^T\right] \sigma_3, 
\nonumber \\
&\quad &^T\Psi^{X,f\dagger}(x) &= - \sigma_3 \left[\Psi^{X,f'}(Tx)^T\right]
(i \sigma_2)^\dagger \tau(Tx)^\dagger, 
\nonumber \\
T':&\quad &^{T'}\Psi^{X,f}(x) &= - \left[\Psi^{X,f'\dagger}(Tx)^T\right] \sigma_3, 
\nonumber \\
  &\quad &^{T'}\Psi^{X,f\dagger}(x) &= \sigma_3 \left[\Psi^{X,f'}(Tx)^T\right].
\end{alignat}
For the Grassmann-valued components we read off
\begin{alignat}{2}\label{trafofermioniccomponentssubmom}
SU(2)_s:&\quad &\psi^{X,f}_\pm(x)' &= \exp(\pm i \alpha(x)) \psi^{X,f}_\pm(x),
\nonumber \\
U(1)_Q:&\quad &^Q\psi^{X,f}_\pm(x) &= \exp(i \omega) \psi^{X,f}_\pm(x),
\nonumber \\
D_i:&\quad &^{D_i}\psi^{X,f}_\pm(x) &= \exp(ik^{f}a_{i})\psi^{X,f}_\pm(x),
\nonumber \\
O:&\quad &^O\psi^{A,\alpha}_\pm(x) &= \mp \exp(-i\tfrac{2\pi}{3})
\exp(\mp i\varphi(Ox))\psi^{B,\beta}_\mp(Ox), \nonumber \\
&\quad &^O\psi^{A,\beta}_\pm(x)&= \mp \exp(i\tfrac{2\pi}{3})
\exp(\mp i\varphi(Ox))\psi^{B,\alpha}_\mp(Ox), \nonumber \\
&\quad &^O\psi^{B,\alpha}_\pm(x) &= \mp \exp(i\tfrac{2\pi}{3})
\exp(\mp i\varphi(Ox))\psi^{A,\beta}_\mp(Ox), \nonumber \\
&\quad &^O\psi^{B,\beta}_\pm(x) &= \mp \exp(-i\tfrac{2\pi}{3})
\exp(\mp i\varphi(Ox))\psi^{A,\alpha}_\mp(Ox), \nonumber \\
O':&\quad &^{O'}\psi^{A,\alpha}_\pm(x) &=\pm
\exp(-i\tfrac{2\pi}{3})\psi^{B,\beta}_\mp(Ox), \nonumber \\
&\quad &^{O'}\psi^{A,\beta}_\pm(x)&= \pm 
\exp(i\tfrac{2\pi}{3})\psi^{B,\alpha}_\mp(Ox), \nonumber \\
&\quad &^{O'}\psi^{B,\alpha}_\pm(x) &= \pm
\exp(i\tfrac{2\pi}{3})\psi^{A,\beta}_\mp(Ox), \nonumber \\
&\quad &^{O'}\psi^{B,\beta}_\pm(x) &= \pm
\exp(-i\tfrac{2\pi}{3})\psi^{A,\alpha}_\mp(Ox), \nonumber \\
R:&\quad &^R\psi^{X,f}_\pm(x) &= \psi^{X,f'}_\pm(Rx), \nonumber \\
T:&\quad &^T\psi^{X,f}_\pm(x) &=
\exp(\mp i \varphi(Tx)) \psi^{X,f'\dagger}_\pm(Tx), \nonumber \\
&\quad &^T\psi^{X,f\dagger}_\pm(x) &= - \exp(\pm i \varphi(Tx))
\psi^{X,f'}_\pm(Tx),
\nonumber \\
T':&\quad &^{T'}\psi^{X,f}_\pm(x) &= - \psi^{X,f'\dagger}_\pm(Tx), \nonumber \\
&\quad &^{T'}\psi^{X,f\dagger}_\pm(x) &= \psi^{X,f'}_\pm(Tx).
\end{alignat}\\
At the moment, the matrix-valued fermion fields have a well-defined
transformation property under $SU(2)_{Q}$. Therefore these fields represent
both  electrons and holes. Since we want to construct an effective theory for
the $t$-$J$ model which contains holes only, a crucial step is to identify the
degrees of freedom that correspond to the holes. In order to remove the
electron degrees of freedom one has to explicitly break the particle-hole
$SU(2)_Q$ symmetry, leaving the ordinary fermion number symmetry $U(1)_Q$
intact. This task can be achieved by constructing all possible fermion mass
terms that are invariant under the various symmetries. Picking the
eigenvectors which correspond to the lowest eigenvalues of the mass matrices
then allows one to separate electrons from holes. The most general mass terms
read
\begin{align}
\sum_{f=\alpha,\beta} & \, \frac{1}{2} \,
\mbox{Tr} \big[ {\cal M} (\Psi^{A,f\dagger} \sigma_3 \Psi^{A,f} -
\Psi^{B,f\dagger} \sigma_3 \Psi^{B,f}) +
m (\Psi^{A,f\dagger} \Psi^{A,f} \sigma_3 +
\Psi^{B,f\dagger} \Psi^{B,f} \sigma_3) \big] \nonumber \\
= &\, \sum_{f=\alpha,\beta} \big[
{\cal M} \big(\psi^{A,f\dagger}_+ \psi^{A,f}_+ - 
\psi^{A,f\dagger}_- \psi^{A,f}_- +
\psi^{B,f\dagger}_- \psi^{B,f}_- - 
\psi^{B,f\dagger}_+ \psi^{B,f}_+ \big) \nonumber \\
&\,  \hspace{2.5em} + m \big(\psi^{A,f\dagger}_+ \psi^{A,f}_+ + 
\psi^{A,f\dagger}_- \psi^{A,f}_- + \psi^{B,f\dagger}_+ \psi^{B,f}_+ + 
\psi^{B,f\dagger}_- \psi^{B,f}_- \big) \big] \nonumber \\
= &\, \sum_{f=\alpha,\beta} \bigg[
\big(\psi^{A,f\dagger}_+, \, \psi^{B,f\dagger}_+ \big) 
\bigg(\begin{array}{cc} {\cal M} + m & 0 \\ 0 & -{\cal M} + m \end{array}\bigg)
\bigg(\begin{array}{c} \psi^{A,f}_+ \\ \psi^{B,f}_+ \end{array}\bigg) 
\nonumber \\
&\, \hspace{2.5em} + \big(\psi^{A,f\dagger}_-, \, \psi^{B,f\dagger}_- \big)
\bigg(\begin{array}{cc} -{\cal M} + m & 0 \\ 0 & {\cal M} + m  \end{array}\bigg)
\bigg(\begin{array}{c} \psi^{A,f}_- \\ \psi^{B,f}_- \end{array}\bigg)
\bigg].
\end{align}
The terms proportional to $\mathcal M$ are invariant under $SU(2)_{Q}$ while
the terms proportional to $m$ are invariant only under the $U(1)_{Q}$ fermion
number symmetry. Since these matrices are already diagonal, we can
directly read off the eigenvalues which are given by $\pm\mathcal M+m$. For
$m=0$ we have a particle-hole symmetric situation. The eigenvalue $\mathcal M$
corresponds to the rest mass of the electrons, while the rest mass of the
holes is given by the eigenvalue $-\mathcal M$. The masses are shifted to
$\pm\mathcal M+m$ when we allow the $SU(2)_{Q}$ breaking term ($m\neq 0$),
which implies that the particle-hole symmetry is destroyed. Hole fields now
correspond to the lower eigenvalue $-\mathcal M+m$ and are identified by the
corresponding eigenvectors $\psi^{B,\alpha}_+(x)$, $\psi^{B,\beta}_+(x)$,
$\psi^{A,\alpha}_-(x)$, and $\psi^{A,\beta}_-(x)$. One can show that these hole
fields and their conjugated counterparts form a closed set under the various
symmetry transformations. We can thus simplify the notation, since a hole
with spin $+$ ($-$) is always located on sublattice $B$ ($A$). Hence, we drop
the sublattice index and the full set of independent low-energy degrees of
freedom describing a doped hole in an antiferromagnet on the honeycomb
lattice is then given by
\begin{alignat}{4}\label{fhl}
\psi_{+}^{\alpha}(x)&=\psi_{+}^{B,\alpha}(x),& \quad\psi_{+}^{\beta}(x)&=
\psi_{+}^{B,\beta}(x),&\quad\psi_{-}^{\alpha}(x)&=\psi_{-}^{A,\alpha}(x),& 
\psi_{-}^{\beta}(x)&=\psi_{-}^{A,\beta}(x),\quad \nonumber \\
\psi_{+}^{\alpha\dagger}(x)&=\psi_{+}^{B,\alpha\dagger}(x),&\quad
\psi_{+}^{\beta\dagger}(x)&=\psi_{+}^{B,\beta\dagger}(x),&\quad
\psi_{-}^{\alpha\dagger}(x)&=\psi_{-}^{A,\alpha\dagger}(x),&\quad
\psi_{-}^{\beta\dagger}(x)&=\psi_{-}^{A,\beta\dagger}(x).
\end{alignat}
Even though $SU(2)_{Q}$ will now no longer be considered as a symmetry of the 
effective theory, it was of central importance for the correct identification 
of the fields for doped holes.

Under the symmetries of the $t$-$J$ model the hole fields transform as
\begin{alignat}{2}\label{finalholefieldtrafos}
SU(2)_s:&\quad &\psi^{f}_\pm(x)' &= \exp(\pm i \alpha(x)) \psi^{f}_\pm(x),
\nonumber \\
U(1)_Q:&\quad &^Q\psi^{f}_\pm(x) &= \exp(i \omega) \psi^{f}_\pm(x),
\nonumber \\
D_i:&\quad &^{D_i}\psi^{f}_\pm(x) &= \exp(ik^{f}a_{i})\psi^{f}_\pm(x),
\nonumber \\
O:&\quad &^O\psi^{\alpha}_\pm(x) &= \mp
\exp(\pm i\tfrac{2\pi}{3}\mp i\varphi(Ox))\psi^{\beta}_{\mp}(Ox), \nonumber \\
&\quad &^O\psi^{\beta}_\pm(x)&=\mp \exp(\mp i\tfrac{2\pi}{3}\mp i \varphi(Ox))
\psi^{\alpha}_{\mp}(Ox), \nonumber \\
O':&\quad &^{O'}\psi^{\alpha}_\pm(x) &= \pm \exp(\pm i\tfrac{2\pi}{3})
\psi^{\beta}_{\mp}(Ox), \nonumber \\
&\quad &^{O'}\psi^{\beta}_\pm(x)&=\pm \exp(\mp i\tfrac{2\pi}{3})
\psi^{\alpha}_\mp(Ox), \nonumber \\
R:&\quad &^R\psi^{f}_\pm(x) &= \psi^{f'}_\pm(Rx), \nonumber \\
T:&\quad &^T\psi^{f}_\pm(x) &= \exp(\mp i \varphi(Tx)) \psi^{f'\dagger}_\pm(Tx),
\nonumber \\
&\quad &^T\psi^{f\dagger}_\pm(x) &= - \exp(\pm i \varphi(Tx)) \psi^{f'}_\pm(Tx),
\nonumber \\
T':&\quad &^{T'}\psi^{f}_\pm(x) &= - \psi^{f'\dagger}_\pm(Tx), \nonumber \\
&\quad &^{T'}\psi^{f\dagger}_\pm(x) &= \psi^{f'}_\pm(Tx).
\end{alignat}
The action to be constructed below must be invariant under all these
symmetries.

\subsection{Low-Energy Effective Lagrangian for Magnons and Holes}

The terms in the action can be characterized by the number $n_\psi$ of fermion
fields they contain, i.e.
\begin{equation}\label{effectiveaction}
S\left[\psi^{f\dagger}_\pm,\psi^f_\pm,v_\mu^\pm,v_\mu^3\right] = \int d^2x \ dt \
\sum_{n_\psi} {\cal L}_{n_\psi}.
\end{equation} 
The leading terms in the effective Lagrangian without fermion fields describe
the pure magnon sector and take the form
\begin{equation}\label{Lagrangian0} 
{\cal L}_0\, = 2\rho_s \left(v^+_i v^-_i + \frac{1}{c^2} v_t^+ v_t^-\right).
\end{equation}
The leading terms with two fermion fields (containing at most one temporal or
two spatial derivatives), describing the propagation of holes as well as their
couplings to magnons, are given by
\begin{align}\label{Lagrangian2}
{\cal L}_2\,=\sum_{\ontopof{f=\alpha,\beta}{\, s = +,-}}\Big[
& M \psi^{f\dagger}_s \psi^f_s + \psi^{f\dagger}_s D_t \psi^f_s
+\frac{1}{2 M'} D_i \, \psi^{f\dagger}_s D_i \psi^f_s 
+\Lambda \psi^{f\dagger}_s (i s v^s_1 + \sigma_f v^s_2) \psi^f_{-s} \nonumber \\
&+iK\big[(D_1 + i s \sigma_f D_2) \psi^{f\dagger}_s
(v^s_1 + i s \sigma_f v^s_2)\psi^f_{-s} \nonumber \\
&\qquad-(v^s_1 + i s \sigma_f v^s_2)\psi^{f\dagger}_s (D_1 + i s  
\sigma_f D_2) \psi^f_{-s} \big] \nonumber \\
&+\sigma_f  L \psi^{f\dagger}_s \epsilon_{ij}f^3_{ij}\psi^f_s 
+ N_1 \psi^{f\dagger}_s v^s_i v^{-s}_i \psi^f_s \nonumber \\
&+ i s \sigma_f N_2 \big( \psi^{f\dagger}_s v^s_1 v^{-s}_2  
\psi^f_s -
\psi^{f\dagger}_s v^s_2 v^{-s}_1 \psi^f_s \big) \Big].
\end{align}
Here $M$ is the rest mass and $M'$ is the kinetic mass of a hole, $\Lambda$
and $K$ are hole-one-magnon couplings, while $L$, $N_1$, and $N_2$ are
hole-two-magnon couplings. Note that all low-energy constants are real-valued.
The sign $\sigma_f$ is $+$ for $\alpha$ and $-$ for $\beta$. We have
introduced the field strength tensor of the composite Abelian ``gauge'' field
\begin{equation}
\label{fstrength}
f^3_{ij}(x) = \p_i v^3_j(x) - \p_j v^3_i(x),
\end{equation}
and the covariant derivatives $D_t$ and $D_i$ acting on $\psi^{f}_{\pm}(x)$ as
\begin{align}\label{kovardrhole}
D_t \psi^f_\pm(x) & = \left[\p_t \pm i v_t^3(x) - \mu \right] \psi^f_\pm(x),
\nonumber \\
D_i \psi^f_\pm(x) &  = \left[\p_i \pm i v_i^3(x)\right] \psi^f_\pm(x).
\end{align}
The chemical potential $\mu$ enters the covariant time-derivative like an
imaginary constant vector potential for the fermion number symmetry $U(1)_Q$.
It is remarkable that the term proportional to $\Lambda$ with just a single
(uncontracted) spatial derivative satisfies all symmetries. Due to the small
number of derivatives it contains, this term dominates the low-energy dynamics
of a lightly hole-doped antiferromagnet on the honeycomb lattice.
Interestingly, for antiferromagnets on the square lattice, a corresponding
term, which was first identified by Shraiman and Siggia, is also present in the 
hole-doped case \cite{Bru06}. On the other hand, a similar term is forbidden
by symmetry reasons in the electron-doped case \cite{Bru07a}. For the honeycomb
geometry we even identify a second hole-one-magnon coupling, $K$, whose
contribution, however, is sub-leading. Interestingly, the field-strength 
tensor $f_{ij}$ appearing in eq.~(\ref{Lagrangian2}) and defined by 
eq.~(\ref{fstrength}) is not allowed for hole- or electron-doped 
antiferromagnets on the square lattice due to symmetry constraints.

The dispersion relation for a single free hole of both flavor $\alpha$ and
$\beta$ can be derived from
$\mathcal L_{2}$ and is given by
\begin{equation}\label{disprelfreehole}
E^{\alpha,\beta}(p) = M + \frac{p_i^2}{2 M'}+\mathcal O(p^{4}),
\end{equation}
which is just the usual dispersion relation for a free non-relativistic
particle. Note that \mbox{$p=(p_{1},p_{2})$} is defined relative to the center
of the hole pockets. Eq.(\ref{disprelfreehole}) confirms that the two pockets
$\alpha$ and $\beta$ are of circular shape which is in agreement with the
result of simulating the one-hole sector of the $t$-$J$ model on the
honeycomb lattice \cite{Jia08}.

The leading terms without derivatives and with four fermion fields are given by
\begin{align}\label{Lagrange4}
{\cal L}_4\,=\sum_{s = +,-} \Big\{
&\frac{G_1}{2} (\psi^{\alpha\dagger}_s \psi^\alpha_s 
\psi^{\alpha\dagger}_{-s} \psi^\alpha_{-s} + 
\psi^{\beta\dagger}_s \psi^\beta_s 
\psi^{\beta\dagger}_{-s} \psi^\beta_{-s}) \nonumber \\
&+G_2 \psi^{\alpha\dagger}_s \psi^\alpha_s \psi^{\beta\dagger}_s \psi^\beta_s 
+ G_3 \psi^{\alpha\dagger}_s \psi^\alpha_s 
\psi^{\beta\dagger}_{-s} \psi^\beta_{-s}\Big\}.
\end{align}
The low-energy four fermion coupling constants $G_{1}$, $G_{2}$, and $G_{3}$
again are real-valued. Although potentially invariant under all symmetries,
terms with two identical hole fields vanish due to the Pauli principle.

\subsection{Accidental Symmetries}

Interestingly, the leading order terms in the effective Lagrangian for magnons
and holes constructed above feature two accidental global symmetries. First,
we notice that for $c\rightarrow\infty$ and without the term proportional to
$iK$ in $\mathcal L_{2}$, Eq.(\ref{Lagrangian0}), Eq.(\ref{Lagrangian2}),
and Eq.(\ref{Lagrange4}) have an accidental Galilean boost symmetry. This
symmetry acts on the magnon and hole fields as
\begin{alignat}{2}\label{boost}
G:&\quad &^GP(x) &= P(Gx), \qquad Gx=(x_{1}-v_{1}t,x_{2}-v_{2}t,t), \nonumber \\
&\quad &^G \psi^f_\pm(x) &= 
\exp(-p^f_{i}x_{i} + \omega^f t) \psi^f_\pm(Gx), \nonumber \\
&\quad &^G \psi^{f\dagger}_\pm(x) &= 
\exp(p^f_{i}x_{i} - \omega^f t) \psi^{f\dagger}_\pm(Gx), \nonumber \\
&\quad &^{G}v_{i}^{3}(x)&=v_{i}^{3}(Gx), \nonumber \\
&\quad &^{G}v_{t}^{3}(x)&=v_{t}^{3}(Gx)-v_{i}v_{i}^{3}(Gx), \nonumber \\
&\quad &^{G}v_{i}^{\pm}(x)&=v_{i}^{\pm}(Gx), \nonumber \\
&\quad &^{G}v_{t}^{\pm}(x)&=v_{t}^{\pm}(Gx)-v_{i}v_{i}^{\pm}(Gx),
\end{alignat}
with 
\begin{align}
\label{pv}
p^f_1=M'v_{1}, \qquad p^f_2=M'v_{2}, \qquad 
\omega^{f}=\frac{(p_{i}^{f})^{2}}{2M'}.
\end{align}
The Galilean boost velocity $\vec{v}$ can be derived alternatively by means of
the hole dispersion relation in Eq.(\ref{disprelfreehole}) and is given by
\mbox{$v_i = d E^f/d p_i^f$} for $i \in \{1,2\}$. Although the Galilean boost symmetry is
explicitly broken at higher orders of the derivative expansion, this symmetry
has physical implications, namely the leading one-magnon exchange between two
holes, to be discussed in the next section, can be investigated in their rest
frame without loss of generality.

In addition, we notice an accidental global rotation symmetry $O(\gamma)$.
Except for the term proportional to $iK$, $\mathcal L_{2}$ of
Eq.(\ref{Lagrangian2}) as well as $\mathcal L_{4}$ of Eq.(\ref{Lagrange4})
are invariant under a continuous spatial rotation by an angle $\gamma$. The
involved fields transform under $O(\gamma)$ as
\begin{eqnarray}
&&^{O(\gamma)}\psi^f_s(x) = \exp(is\sigma_{f} \tfrac{\gamma}{2})
\psi^f_s(O(\gamma)x), \quad s = \pm, \nonumber \\
&&^{O(\gamma)} v_1(x) = \cos\gamma \ v_1(O(\gamma)x) + \sin\gamma \ 
v_2(O(\gamma)x), \nonumber \\
&&^{O(\gamma)} v_2(x) = - \sin\gamma \ v_1(O(\gamma)x) + \cos\gamma \ 
v_2(O(\gamma)x),
\end{eqnarray}
with
\begin{eqnarray}
&&O(\gamma)x = O(\gamma)(x_1,x_2,t) = (\cos\gamma \ x_1 - \sin\gamma \ x_2,\sin
\gamma \ x_1 + \cos\gamma \ x_2,t).
\end{eqnarray}
Here $v_i$ denotes the composite magnon field. This symmetry is not present
in the $\Lambda$-term of the square lattice. The $O(\gamma)$ invariance has
some interesting implications for the spiral phases in a lightly doped
antiferromagnet on the honeycomb lattice and was investigated in detail in 
\cite{Jia09}. 

\section{One-Magnon Exchange Potentials}
\label{potential}

In the effective theory framework, at low energies, holes interact with each
other via magnon exchange. Since the long-range dynamics is dominated by
one-magnon exchange, we will calculate the one-magnon exchange potentials
between two holes of the same flavor $\alpha$ and $\beta$ and of different
flavor. 

\begin{figure}[t]
\begin{center}
\vspace{-0.4cm}
\includegraphics[width=7cm]{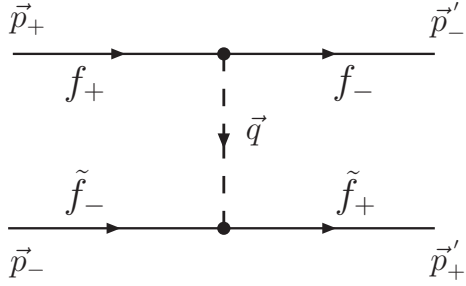}
\end{center}
\caption{\it Tree-level Feynman diagram for one-magnon exchange between two
holes.}
\label{fig_Feynman}
\end{figure}

In order to address the one-magnon physics, we expand in the magnon
fluctuations $m_1(x)$ and $m_2(x)$ around the ordered staggered magnetization
\begin{equation}
\vec e(x) = \left( \frac{m_1(x)}{\sqrt{\rho_s}},\,
\frac{m_2(x)}{\sqrt{\rho_s}},1 \right) + {\cal O}\left(m^2\right).
\end{equation}
For the composite magnon fields this leads to
\begin{align}
v_\mu^\pm(x) &= \frac{1}{2 \sqrt{\rho_s}} \p_\mu
\big[ m_2(x) \pm i m_1(x) \big] + {\cal O}\left(m^3\right), \nonumber \\
v_\mu^3(x) &= \frac{1}{4 \rho_s}\big[m_1(x) \p_\mu m_2(x) -
m_2(x) \p_\mu m_1(x)\big] + {\cal O}\left(m^4\right).
\end{align}
Since vertices with $v_\mu^3(x)$ involve at least two magnons, one-magnon
exchange results from vertices with $v_\mu^\pm(x)$ only. As a consequence, two
holes can exchange a single magnon only if they have anti-parallel spins ($+$
and $-$), which are both flipped in the magnon-exchange process. We denote the
momenta of the incoming and outgoing holes by $\vec p_\pm$ and $\vec p_\pm\!'$,
respectively. The momentum carried by the exchanged magnon is denoted by
$\vec q$. The incoming and outgoing holes are asymptotic free particles with
momentum $\vec p=(p_1,p_2)$ and energy $E(\vec p)= M +  p_i^2/2M'$. One-magnon
exchange between two holes is associated with the Feynman diagram in
Figure~\ref{fig_Feynman}.
Evaluating these Feynman diagrams, in momentum space one arrives at the following 
potentials for various combinations of flavors $f, \tilde f \in \{\alpha, \beta\}$ 
and couplings $F,\tilde F \in\{\Lambda,K\}$ 
\begin{equation}
\langle \vec p_+ \ \!\!\!\! '\, \vec p_- \ \!\!\!\! '|V^{f\tilde f}_{F\tilde F}
|\vec p_+ \vec p_-\rangle = V^{f\tilde f}_{F\tilde F}(\vec q \,) \
\delta(\vec p_+ + \vec p_- - \vec p_+ \ \!\!\!\! ' - \vec p_- \ \!\!\!\! '),
\qquad F,\tilde F \in \left\{\Lambda,K\right\},
\end{equation}
with
\begin{align}
V^{ff}_{\Lambda \Lambda}(q) &= - \frac{\Lambda^2}{2\rho_s},\qquad
V^{ff'}_{\Lambda \Lambda}(q) = \frac{\Lambda^2}{2\rho_sq^2}\big(iq_1-\sigma_fq_2
\big)^2, \nonumber \\
V^{ff}_{KK}(q) &= - \frac{K^2}{2\rho_s}\big[2(p_{+1}-i\sigma_fp_{+2})-q_1
+i\sigma_fq_2\big]\big[2(p_{-1}+i\sigma_fp_{-2})+q_1+i\sigma_fq_2\big],
\nonumber \\
V^{ff'}_{KK}(q) &= - \frac{K^2}{2\rho_sq^2}\big(q_1-i\sigma_f q_2\big)^2
\big[2(p_{+1}-i\sigma_fp_{+2})-q_1+i\sigma_fq_2\big] \nonumber \\
&\times\big[2(p_{-1}-i\sigma_fp_{-2})+q_1-i\sigma_fq_2\big], \nonumber \\
V^{ff}_{\Lambda K}(q) &= - \frac{i\Lambda K}{2\rho_sq^2}
\big(q_1+i\sigma_fq_2\big)^2
\big[2(p_{-1}+i\sigma_fp_{-2})+q_1+i\sigma_fq_2\big], \nonumber \\
V^{ff'}_{\Lambda K}(q) &= - \frac{i\Lambda K}{2\rho_s}
\big[2(p_{-1}-i\sigma_fp_{-2})+q_1-i\sigma_fq_2\big], \nonumber \\
V^{ff}_{K \Lambda}(q) &= \frac{iK \Lambda}{2\rho_sq^2}
\big(q_1-i\sigma_fq_2\big)^2
\big[2(p_{+1}-i\sigma_fp_{+2})-q_1+i\sigma_fq_2\big], \nonumber \\
V^{ff'}_{K \Lambda}(q) &= \frac{iK \Lambda}{2\rho_s}
\big[2(p_{+1}-i\sigma_fp_{+2})-q_1+i\sigma_fq_2\big].
\end{align}
We noted earlier that the leading contribution to the low-energy physics comes
from the $\Lambda$-vertex. From here on, we therefore concentrate on the
potential with two $\Lambda$ vertices only. In coordinate space the
$\Lambda\Lambda$-potentials are given by
\begin{equation}\label{general_potentials}
\langle \vec r_+ \ \!\!\!\! ' \vec r_- \ \!\!\!\! '|V^{f\tilde f}_{\Lambda\Lambda}
|\vec r_+ \vec r_-\rangle = V^{f\tilde f}_{\Lambda\Lambda}(\vec r\, ) \
\delta(\vec r_+ - \vec r_- \ \!\!\!\! ') \
\delta(\vec r_- - \vec r_+ \ \!\!\!\! '),
\end{equation}
with
\begin{equation}
V^{ff}_{\Lambda\Lambda}(\vec r\,) = -\frac{\Lambda^2}{2\rho_s}\delta^{(2)}(\vec r
\,),\qquad
V^{ff'}_{\Lambda\Lambda}(\vec r\,) = \frac{\Lambda^2}{2\pi \rho_s\vec r^{\,2}}
\exp (2i\sigma_f \varphi).
\end{equation}
Here $\vec r = \vec r_+ - \vec r_-$ denotes the distance vector between the
two holes and $\varphi$ is the angle between $\vec r$ and the $x_1$-axis. The
\mbox{$\delta$-f}unctions in Eq.(\ref{general_potentials}) ensure that the
holes do not change their position during the magnon exchange. It should be
noted that the one-magnon exchange potentials are instantaneous although
magnons travel with the finite speed c. Retardation effects occur only at
higher orders. 

Interestingly, in the $\Lambda\Lambda$ channel, one-magnon exchange over long
distances between two holes can only happen for holes of opposite flavor. For
two holes of the same flavor, one-magnon exchange acts as a contact
interaction. In the next section we will concentrate on the long-range physics
of weakly bound states of holes and therefore we will only consider the binding
of holes of different flavor. 

\section{Two-Hole Bound States}
\label{boundstates}

We now investigate the Schr\"odinger equation for the relative motion of two
holes with flavors $\alpha$ and $\beta$. In the following, we will treat short 
distance interactions by imposing a hard-core boundary condition on the pair's 
wave function. Due to the accidental Galilean boost invariance,
without loss of generality, we can consider the hole pair in its rest frame.
The total kinetic energy of the two holes is given by
\begin{equation}
T =  \sum_{f=\alpha,\beta} T^{f} = \sum_{f = \alpha,\beta} \frac{p_i^2}{2 M'}
= \frac{p_i^2}{M'}. 
\end{equation}
We introduce the two probability amplitudes \mbox{$\Psi_1(\vec r \,)$} and
\mbox{$\Psi_2(\vec r \,)$} which represent the two flavor-spin combinations
\mbox{$\alpha_+\beta_-$} and \mbox{$\alpha_-\beta_+$}, respectively, where we
choose the distance vector \mbox{$\vec r$} to point from the \mbox{$\beta$} to
the \mbox{$\alpha$} hole. Since the holes undergo a spin flip during the
magnon exchange, the two probability amplitudes are coupled through the magnon
exchange potentials and the Schr\"odinger equation describing the relative
motion of the hole pair is a two-component equation. Using the explicit form
of the potentials, the relevant Schr\"odinger equation for two holes reads
\begin{equation}
\left(\begin{array}{cc} - \frac{1}{M'}\Delta & \gamma\frac{1}{\vec r^{\,2}}
\exp(-2i\varphi)
\\[0.2ex]
\gamma\frac{1}{\vec r^{\,2}}\exp(2i\varphi) &  - \frac{1}{M'} \Delta 
\end{array} \right)
\left(\begin{array}{c} \Psi_1(\vec r \, ) \\ 
\Psi_2(\vec r \, ) \end{array}\right) = E 
\left(\begin{array}{c} \Psi_1(\vec r \, ) \\ 
\Psi_2(\vec r \, ) \end{array}\right),
\end{equation}
with
\begin{equation}\label{gamma}
\gamma = \frac{\Lambda^2}{2\pi \rho_s}.
\end{equation}
Making the separation ansatz
\begin{equation}
\label{ansatz}
\Psi_1(r,\varphi) = R_1(r) \exp (i m_1 \varphi), \qquad \Psi_2(r,\varphi) =
R_2(r) \exp (i m_2 \varphi),
\end{equation}
with \mbox{$r= \left|\vec r\,\right|$}, and using the Laplace operator in
polar coordinates one arrives at the coupled equations
\begin{align}
-\left(\frac{d^2}{d r^2}+\frac{1}{r}\frac{d}{d r} -\frac{1}{r^2}m_1^2 \right)
R_1(r)+\gamma M'\frac{R_2(r)}{r^{2}}\exp \big(-i\varphi(2+m_1-m_2)\big) &= M' E
 R_1(r), \nonumber\\
-\left(\frac{d^2}{d r^2}+\frac{1}{r}\frac{d}{d r} -\frac{1}{r^2}m_2^2 \right)
R_2(r)+\gamma M'\frac{R_1(r)}{r^{2}}\exp \big(i\varphi(2+m_1-m_2)\big) &= M' E 
R_2(r).
\end{align}
The radial and angular part can be separated provided that the condition 
$m_2 - m_1 = 2$ is satisfied. Introducing the parameter $m$, which is implicitly
defined by
\begin{equation}
m_1=m-1, \qquad m_2=m+1,
\end{equation}
the radial equations are then given by
\begin{align}\label{radial_equations}
-\left(\frac{d^2}{d r^2}+\frac{1}{r}\frac{d}{d r} -\frac{1}{r^2}(m-1)^2 \right)
R_1(r)+\gamma M'\frac{R_2(r)}{r^{2}} &= M' E R_1(r), \nonumber\\
-\left(\frac{d^2}{d r^2}+\frac{1}{r}\frac{d}{d r} -\frac{1}{r^2}(m+1)^2 \right)
R_2(r)+\gamma M'\frac{R_1(r)}{r^{2}} &= M' E R_2(r).
\end{align}
While the other cases would have to be investigated numerically, for $m=0$ the 
two radial equations decouple and can be solved analytically. In particular, 
by taking appropriate linear combinations, for $m=0$ the two equations can be 
cast into the form
\begin{align}
\left[-\left(\frac{d^2}{d r^2}+\frac{1}{r}\frac{d}{d r}\right) +(1 +\gamma M')
\frac{1}{r^{2}}\right]\big(R_1(r)+R_2(r)\big) &= M' E \big(R_1(r)+R_2(r)\big),
\nonumber\\
\left[-\left(\frac{d^2}{d r^2}+\frac{1}{r}\frac{d}{d r}\right) +(1 -\gamma M')
\frac{1}{r^{2}}\right]\big(R_1(r)-R_2(r)\big) &= M' E \big(R_1(r)-R_2(r)\big).
\end{align}
Because the two equations are different, but contain the same energy $E$, one
of the equations has a vanishing solution. In the first equation the potential
always has a positive sign and is thus repulsive. In the second equation, on
the other hand, the potential has a negative sign and is therefore attractive
when the low-energy constants obey the relation
\begin{equation}
1 - \gamma M' = 1 - \frac{M'\Lambda^2}{2\pi \rho_s} \leq 0.
\end{equation}
Thus, magnon-mediated forces can lead to bound states only if the low-energy
constant $\Lambda$ is larger than the critical value
\begin{equation}
\Lambda_c = \sqrt{\frac{2\pi \rho_s}{M'}}.
\end{equation}
Interestingly, the same critical value arises in the investigation of spiral
phases in a lightly doped antiferromagnet on the honeycomb
lattice \cite{Jia09}. There it marks the point where spiral phases become
energetically favorable compared to the homogeneous phase. Here we are
interested in the solution of the above system where the first equation has a
zero solution and the second a non-zero one, i.e.\ $R_1(r)+R_2(r)=0$.
Identifying $R(r)=R_1(r)-R_2(r)$, the second equation takes the form
\begin{equation}
\label{radialeq}
\left[-\left(\frac{d^2}{d r^2}+\frac{1}{r}\frac{d}{d r}\right) +(1 -\gamma M')
\frac{1}{r^{2}}\right]R(r) = - M' |E| R(r),
\end{equation}
where we have set $E=-|E|$.
The same equation occurred in the square lattice case \cite{Bru06,Bru06a}
and can be solved along the same lines. As it stands, the equation is
ill-defined because the $1/r^2$ potential is too singular at the origin.
However, we have not yet included the contact interaction proportional to the
4-fermion coupling $G_3$. Here, in order to keep the calculation analytically
feasible, we model the short-range repulsion by a hard core radius $r_0$, i.e.\
we require \mbox{$R(r_0)=0$} for \mbox{$r \leq r_0$}. Eq.(\ref{radialeq}) is
solved by a modified Bessel function
\begin{equation}
R(r) = A K_\nu \big( \sqrt{M' |E|} r \big), \qquad \nu = i \sqrt{\gamma M' -1},
\end{equation}
with $A$ being a normalization constant. Demanding that the wave function
vanishes at the hard core radius gives a quantization condition for the bound
state energy. The quantum number $n$ then labels the $n$-th excited state. For
large $n$, the binding energy is given by
\begin{equation}
E_n \sim - \frac{1}{M' r_0^2} \exp\left(\frac{- 2 \pi n}{\sqrt{\gamma M' -1}}
\right).
\end{equation}
Like every quantity calculated within the framework of the effective theory,
the binding energy depends on the values of the low-energy constants. The 
binding is exponentially small in $n$ and there are infinitely many bound
states. While the highly excited states have exponentially small energy, for
sufficiently small $r_0$ the ground state could have a small size and be
strongly bound. However, as already mentioned, for short-distance physics the
effective theory should not be trusted quantitatively. If the holes were
really tightly bound, one could construct an effective theory which
incorporates them explicitly as relevant low-energy degrees of freedom. As
long as the binding energy is small compared to the relevant high-energy
scales, our result is valid and receives only small corrections from
higher-order effects such as two-magnon exchange.

\begin{figure}[t]
\begin{center}
\vspace{-0.4cm}
\includegraphics[width=7cm]{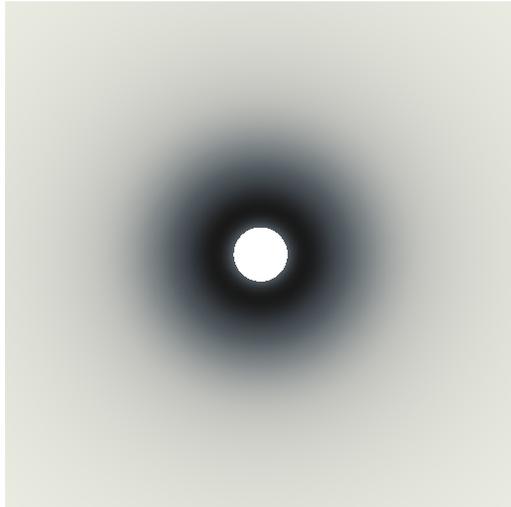}
\end{center}
\caption{\it Probability distribution for the ground state of two holes of
flavors $\alpha$ and $\beta$.}
\label{distribution}
\end{figure}

Finally, let us discuss the angular part of the wave equation. The ansatz 
(\ref{ansatz}) leads to the following solution for the ground state wave 
function
\begin{equation}
\label{groundstatewavefunction}
\Psi(r, \varphi) = 
\left(\begin{array}{c} \Psi_1(\vec r \, ) \\ 
\Psi_2(\vec r \, ) \end{array}\right)
= R(r) \, 
\left(\begin{array}{c} \exp(- i \varphi) \\ 
- \exp(i \varphi) \end{array}\right) .
\end{equation}
Applying the 60 degrees rotation $O$ and using the transformation rules of
Eq.(\ref{finalholefieldtrafos}) one obtains
\begin{equation}
^O\Psi(r, \varphi) = - \, \Psi(r, \varphi).
\end{equation}
Interestingly, the wave function for the ground state of two holes of flavors
$\alpha$ and $\beta$ thus exhibits $f$-wave symmetry.\footnote{Strictly 
speaking, the continuum classification scheme of angular momentum eigenstates 
does not apply here, since we are not dealing with a continuous rotation 
symmetry.} The corresponding probability distribution depicted in 
Figure~\ref{distribution}, on the other hand, seems to show $s$-wave symmetry. 
However, the relevant phase information is not visible in this picture, because 
only the probability density is shown.
Interestingly, for two-hole bound states on the square lattice, the wave 
function for the the ground state of two holes of flavors $\alpha$ and $\beta$ 
shows $p$-wave symmetry, while the corresponding probability distribution (which
again does not contain the relevant phase information) resembles $d_{x^2-y^2}$ 
symmetry \cite{Bru06}. Remarkably, the ground state wave function 
(\ref{groundstatewavefunction}) of a bound hole pair on the honeycomb lattice 
remains invariant under the reflection symmetry $R$, the shift symmetries 
$D_i$, as well as under the accidental continuous rotation symmetry $O(\gamma)$.

We would like to emphasize that the $f$-wave character of the two-hole bound 
state on the honeycomb lattice is an immediate consequence of the systematic 
effective field theory analysis. It seems that the issue of the true symmetry 
of the pairing state, realized in the dehydrated version of 
Na$_2$CoO$_2 \times y$H$_2$O is still controversial \cite{Iva09}. Still, it is 
quite interesting to note that a careful analysis of the available experimental
data for this compound suggests that the pairing symmetry indeed is $f$-wave 
\cite{Maz05}.

\section{Conclusions}
\label{conclusions}

In complete analogy to our earlier investigations on the square lattice, we have
constructed a systematic low-energy effective field theory of magnons and doped
holes in an antiferromagnet on the honeycomb lattice. Due to the different 
lattice geometry, there are important symmetry differences which have an
impact on the allowed terms that enter the effective Lagrangian. Interestingly,
in contrast to the square lattice case, on the honeycomb lattice an accidental 
continuous spatial rotation invariance arises for the leading terms of the 
low-energy effective Lagrangian.

As an important result, we have identified the leading magnon-hole vertex which
results from a term with a single uncontracted spatial derivative. This term,
which is analogous to the Shraiman-Siggia term on the square lattice, yields a
rather strong magnon-hole coupling since it appears at a low order in the
systematic low-energy expansion. As we have investigated earlier, at non-zero
hole doping, when $\Lambda$ is sufficiently strong, this term gives rise to 
spiral phases in the staggered magnetization \cite{Jia09}.

In the present work, we have studied the effect of the magnon-hole vertex on
two-hole bound states. Again in contrast to the square lattice case, it turned
out that the magnon-hole coupling constant $\Lambda$ must exceed a critical 
value in order to obtain two-hole bound states. Our analysis implies that the 
wave function for the ground state of two holes of flavors $\alpha$ and $\beta$
exhibits $f$-wave symmetry (while the corresponding probability distribution 
seems to suggest $s$-wave symmetry). This is quite different from the square 
lattice case, where the wave function for the ground state of two holes of 
flavors $\alpha$ and $\beta$ exhibits $p$-wave symmetry (while the 
corresponding probability distribution resembles $d_{x^2-y^2}$ symmetry).

We like to stress again that the effective theory provides a  theoretical 
framework in which the low-energy dynamics of lightly hole-doped antiferromagnets 
can be investigated in a systematic manner. Once the low-energy parameters have 
been adjusted appropriately by comparison with either experimental data or numerical 
simulations, the resulting physics is completely equivalent to the one of the Hubbard 
or $t$-$J$ model.

\section*{Acknowledgments}

F.~K. acknowledges that he has made most of his contributions to this paper
while working in the Condensed Matter Theory Group at the Massachusetts
Institute of Technology. U.-J.\ W.\ likes to thank the members of the Center
for Theoretical Physics at MIT, where part of this work was done, for their
hospitality. C.~P.\ H.\ thanks the Institute for Theoretical Physics at Bern
University for their warm hospitality and gratefully acknowledges financial
support from the Universidad de Colima which made a stay at Bern University
possible. F.-J.\ J.\ is partially supported by NCTS (North) and NSC (Grant
No. NSC 99-2112-M003-015-MY3) of R.~O.~C.. This work was also supported in 
part by funds provided by the Schweizerischer Nationalfonds (SNF). In 
particular, F.\ K.\ was supported by an SNF young researcher fellowship. 
The ``Albert Einstein  Center for Fundamental Physics'' at Bern University 
is supported by the ``Innovations-und Kooperationsprojekt C-13'' of the 
Schweizerische Uni\-ver\-si\-t\"ats\-kon\-fe\-renz (SUK/CRUS).

\end{document}